\documentclass[twoside,12pt]{article}
\usepackage[utf8]{inputenc}
\usepackage{xr-hyper}
\usepackage{hyperref}
\usepackage{url}
\usepackage{xspace}
\usepackage{graphicx}
\usepackage{authblk}
\usepackage{xcolor}
\RequirePackage{lineno}
\usepackage[centering]{geometry}

\newcommand{\pA}{\mbox{$pA$}\xspace}

\newcommand{\auau}{\mbox{Au$+$Au}\xspace} 

\newcommand{\pbpb}{\mbox{Pb$+$Pb}\xspace}
\newcommand{\Raa}{\mbox{$R_{\mathrm{AA}}$}}
\newcommand{\Rcp}{\mbox{$R_{\mathrm{CP}}$}}
\newcommand{\yjet}{\mbox{$y^{\mathrm{jet}}$}}

\newcommand {\pp}{\mbox{$pp$}}

\newcommand{\pT}{\mbox{${p_T}$}\xspace}
\newcommand{\bjet}{\mbox{$b$}-jet\xspace}
\newcommand{\bjets}{\mbox{$b$}-jets\xspace}

 \newcommand{\qhat}{\mbox{$\hat{q}$}}
 
 \newcommand{\pPb}{\mbox{$p$$+$Pb}}
\newcommand{\RTwo}{\mbox{$R=0.2$}}
\newcommand{\RThree}{\mbox{$R=0.3$}}
\newcommand{\z}{\mbox{$z$}}
\newcommand{\RFour}{\mbox{$R=0.4$}}
\newcommand{\xexe}{\mbox{Xe$+$Xe}\xspace} 
\newcommand{\Npart}{\mbox{$N_{part}$}\xspace} 
\newcommand{\OO}{\mbox{O$+$O}\xspace}

\newcommand{\RAA}{\mbox{$R_{AA}$}\xspace} 
\newcommand{\RpA}{\mbox{$R_{pA}$}\xspace} 
\newcommand{\TAA}{\mbox{$\langle T_{AA}\rangle$} \xspace} 
\newcommand{\RpPb}{\mbox{$R_{pPb}$}\xspace} 

\newcommand{\pt}{\mbox{${p_T}$}\xspace}
\newcommand{\ptjet}{\mbox{$p_{\mathrm{T}}^{\mathrm{jet}}$}}
\newcommand{\ptgamma}{\mbox{$p_{\mathrm{T}}^{\mathrm{\gamma}}$}}
\newcommand{\ptpart}{\mbox{$p_{\mathrm{T}}^{\mathrm{part}}$}}
\newcommand{\vn}{\mbox{$v_{n}$}}
\newcommand{\xJ}{\mbox{$x_{J}$}}
\newcommand{\pizero}{\mbox{$\pi^{0}$}}
\newcommand{\xJgamma}{\mbox{$x_{J\gamma}$}}
\newcommand{\Dr}{\mbox{$\Delta R$}}
\newcommand{\vtwo}{\mbox{$v_{\mathrm{2}}$}}
\newcommand{\vthree}{\mbox{$v_{\mathrm{3}}$}}
\newcommand{\vfour}{\mbox{$v_{\mathrm{4}}$}}

\newcommand{\dAu}{\mbox{$d$$+$Au}\xspace}

\newcommand{\pythia}{\mbox{\sc Pythia}\xspace}

\title{Studying the QGP with Jets at the LHC and RHIC}
\author[1]{Leticia Cunqueiro}
\author[2]{Anne M. Sickles}
\affil[1]{\normalsize Laboratoire Leprince-Ringuet, Ecole Polytechnique, Palaiseau, France}
\affil[2]{\normalsize Department of Physics, University of Illinois, Urbana IL, USA}
\date{\today}

\begin{document}

\maketitle
\tableofcontents
\section{Introduction}
The main goal of the heavy-ion physics program at the Large Hadron
Collider (LHC) and the Relativistic Heavy Ion Collider (RHIC) is
to study quantum chromodynamics (QCD) at extremely high
temperature.  In order to do this, heavy nuclei are collided at ultrarelativistic energies.  As the nuclei pass through each other, 
a region of extremely large energy density is created (greater than $12$ GeV/fm$^{3}$ 1 fm after the collision \cite{Busza:2018rrf}) and this results in the creation of  matter known as the quark-gluon plasma (QGP)~\cite{BRAHMS:2004adc,PHENIX:2004vcz,PHOBOS:2004zne,STAR:2005gfr,Roland:2014jsa}.  A major discovery of the RHIC and LHC experimental programs is that the QGP is well described as a 
nearly ideal fluid~\cite{Heinz:2013th} with a maximal temperature of at least 300~MeV~\cite{Adam:2015lda}. This strongly coupled 
fluid exhibits a viscosity to entropy ratio near the conjectured
lower limit of $\frac{1}{4\pi}$~\cite{Kovtun:2004de}, expected for quantum fluids \cite{Adams:2012th} that can be described in a dual gravity picture~\cite{Maldacena:1997re}.

A key aim is to understand how such a strongly correlated liquid arises from the underlying theory, QCD, and its degrees of freedom, the quarks and gluons. 
%This matter is qualitatively different than nuclear matter at lower temperatures.
Jet measurements in heavy-ion collisions are of great interest to study the microscopic structure of the QGP liquid. Since jets are multi-scale objects, they probe the QGP at varying length scales. Jets
have been identified as central to understanding the nature of the interactions
which give rise to the fluid-like behavior of the QGP~\cite{Aprahamian:2015qub,EuropeanStrategyforParticlePhysicsPreparatoryGroup:2019qin}.

Jets in hadronic collisions are formed by the point-like scattering of 
quarks and/or gluons.  
Jets are well-defined objects in QCD and are under good theoretical
and experimental control in \pp\
collisions (see e.g Refs.~\cite{ATLAS:2017ble,CMS:2016jip,Marzani:2019hun}). 
In heavy-ion collisions, the hard, elementary scatterings leading to jet production occur in the early stages of the collision. 
The evolution of the scattered quark or gluon towards hadronisation is 
embedded with and interacts with the evolving QGP medium, and is thus subject to modifications relative to $pp$ collisions.

The first description of the propagation of an energetic parton (quark or gluon) in the QGP appeared in Ref.~\cite{Bjorken:1982tu}.
Further studies identified the dominant mechanism of energy loss for high-energy partons to be gluon radiation induced by the QGP~\cite{Gyulassy:1990ye,Baier:1996kr,Baier:1996sk}.
QGP-induced modifications to jet properties are generically called jet quenching because the most direct consequence of parton energy loss in the QGP is the reduced 
energy of jets, resulting in 
a reduced number of reconstructed hadrons and jets in heavy-ion collisions at a fixed momentum compared to expectations from \pp\ collisions.

Jet measurements in heavy ion collisions, as we will discuss in the next sections, attempt to capture the full dynamics of jet quenching across 
different jet radii, collision
geometry and energy. They comprise survival rates constructed as ratios of jet (or hadrons from jets) cross sections relative to expectations based on \pp \,  collisions as well as 
the jet radius, $R$, dependence of jet cross sections, inter-jet correlations, jet 
azimuthal anisotropies and measurements of the jet shapes, fragmentation and substructure.

Determining QGP properties from the jet modifications is not trivial.  First, the precise mechanisms of jet-medium interactions are currently under investigation and the predictive power of the different theoretical formalisms and approximations are still to be validated. 
Second, jet measurements are often affected by multiple confounding 
effects. Also, any measurement of jets is necessarily made after it has propagated through the entire time evolution of the QGP and effects preceeding and following the QGP existence can impact measured quantities. Finally, jet measurements in heavy-ion collisions are experimentally challenging due to the large and fluctuating background from the underlying event in a typical heavy-ion collision.

%Thirdly, the evolving structure
%of each jet fragmenting itself is unique meaning that different types of 
%jets are expected to interact with the QGP in different ways.

In this review, we will first discuss how jet measurements in heavy-ion collisions are performed.  
Then we will focus on three important questions  related to the physics of jet quenching whose answers are not yet complete but will be within reach in the next few years due to new experimental data from
RHIC and the LHC and theoretical advances:
\begin{itemize}
\item How is jet energy transported within the QGP? 
\item What are the effective degrees of freedom of the QGP?
\item Is there a critical size for QGP formation?
\end{itemize}

We will end with a brief conclusion and outlook.

\section{A brief summary of the theoretical advances in jet quenching} 
A highly energetic parton that propagates through high-temperature and high-density QCD matter is expected to lose energy mainly via radiative 
processes~\cite{Gyulassy:1990ye,Baier:1996kr,Baier:1996sk}.
These processes 
consist of gluon radiation induced by the 
scattering of the energetic parton with the medium constituents.
A radiated single gluon spectrum  master formula was derived within the 
Baier-Dokshitzer-Mueller-Peigne-Schiff-Zakharov  (BDMPS-Z)
formalism in the 1990s~\cite{ Baier:1996sk, Baier:1996kr,Zakharov:1997uu}. Two limiting approximations, allowing for a semi-analytical treatment of the calculations to make them numerically tractable, are traditionally considered.
\begin{itemize} 
\item the limit where the interactions with the medium are few hard scatterings, where an expansion in terms of the number of scatterings with the medium is possible -also known as the opacity expansion, independently derived by Gyulassy-Levai-Vitev (GLV)~\cite{Gyulassy:1999zd}.
\item the limit where the interactions where there are  multiple, coherent and soft interactions with the medium~\cite{Salgado:2003gb,Baier:1996sk, Baier:1996kr,Zakharov:1997uu}. In this limit, the medium-induced gluon spectrum is controlled by a single parameter, the transport coefficient $\hat{q}$, which quantifies the average momentum transferred from the medium to the parton per unit path length. Similar resummation of multiple scatterings are also considered in the Arnold-Moore-Yaffe (AMY) formalism \cite{Arnold:2001ba,Arnold:2001ms,Arnold:2002ja} where the complete thermal propagators can be included at the price of considering only infinitely long media -this limitation was overcome in \cite{Caron-Huot:2010qjx}.
\end{itemize}

The dilute medium approximation, where only a few, normally one, scattering is considered, is also the focus of the "Higher-twist" approach in which induced gluon radiation is computed in the DIS kinematics as a higher-twist radiative correction to the inclusive cross section \cite{Wang:2001ifa,Majumder:2007hx}. 

Further important theoretical developments followed, such as the calculation of the medium-induced gluon spectra off a $q{\overline{q}}$ antenna \cite{Mehtar-Tani:2010ebp,Mehtar-Tani:2011hma,Mehtar-Tani:2011vlz,Casalderrey-Solana:2011ule,Armesto:2011hh,Armesto:2011ir,Mehtar-Tani:2012mfa,Casalderrey-Solana:2012evi} which addressed color coherence in multi-gluon emissions in medium. These developments exposed a new 
transverse scale, the medium correlation length. This scale dictates up to which transverse distance a pair of partons remains color-coherent and thus resolved by the medium as a single color charge. The role of these interferences in the two-gluon radiation spectrum was also studied in a series of papers \cite{Arnold:2015qya,Arnold:2016kek,Arnold:2020uzm}.

Recent theoretical work aims at increasing the precision of the analytical calculations by relaxing the approximations. The main difficulty is the correct treatment of the multiple scatterings with the medium and their interference as described by the Landau-Pomeranchuk-Migdal (LPM) effect~\cite{LPM}.
 This has been done numerically \cite{Andres:2020vxs,Feal:2018sml} or at 
next-to-leading order (NLO) in a new expansion scheme \cite{Barata:2021wuf}, the Improved Opacity Expansion (IOE)~\cite{Mehtar-Tani:2019ygg}.
A medium-induced gluon spectrum and further jet observables that incorporate both the soft and hard limits are essential as a theoretical reference for 
probing the shortest length scales
in the QGP, including answering whether it is possible to resolve point-like scattering
within the QGP. %ongoing experimental searches of point-like scatterers within the QGP. 

In some cases, analytical calculations of jet observables are available. For instance, a first-principle calculation of the jet $R$ dependence of inclusive jet suppression was recently presented \cite{Mehtar-Tani:2021fud}, incorporating both latest NLO calculations of the gluon spectrum in the IOE and color coherence effects. The Soft Collinear effective theory SCET~\cite{Bauer:2000yr} and its extension to heavy ions, SCET$_{G}$~\cite{Idilbi:2008vm,Ovanesyan:2011xy} have provided a framework to calculate jet cross sections and substructure \cite{Chien:2016led,Chien:2015hda}.

In order to do jet phenomenology and to compare theory expectations to jet measurements, Monte Carlo implementations are used in most of the cases since they bring in important higher-order corrections via the parton shower and the possibility to include 
the development of the parton shower through the lifetime of the QGP. Theoretical prescriptions for single gluon medium-induced emissions are incorporated into Monte Carlo generators \cite{Caucal:2020zcz,Zapp:2013vla,Schenke:2009gb,JETSCAPE:2020ttu,Armesto:2008qh,Renk:2010zx,Lokhtin:2011qq}.

As the jet shower develops and the jet constituents become softer and softer, the fate of such energy and the medium response to it can be described using transport 
models~\cite{Wang:2013cia,He:2018xjv,He:2015pra,Schenke:2009gb}.
Some models describe the medium response via recoil particles while others describe it hydrodynamically after
local thermalisation (see Ref.~\cite{Cao:2020wlm} for a review).

All the above developments rely on the applicability of perturbative QCD. However jets are multiscale objects and as the partons in the shower evolve, they will reach scales that are of the order of the QGP temperature for which a weak coupling description might no longer be valid. The dual gravity picture 
has made possible to use holographic calculations to study energy loss in a strongly coupled QGP~\cite{Chesler:2014jva}. The Hybrid Model~\cite{Casalderrey-Solana:2014bpa} uses a Monte 
Carlo approach that incorporates holographic prescriptions for energy loss.

This section is meant to briefly present the theoretical context for the jet measurements in heavy ion collisions that will be discussed in the following. We point the reader to some recent theory reviews~\cite{Qin:2015srf,Blaizot:2015lma,Casalderrey-Solana:2007knd} for further reading.

\section{Jet Measurements in Heavy Ion Collisions}
\subsection{Jet Reconstruction}
The standard jet finding algorithm used in heavy ion collisions is the anti$-k_{\rm{T}}$ 
algorithm~\cite{Cacciari:2008gp} as implemented in the FastJet package~\cite{Cacciari:2011ma} due to
its wide adoption in the high-energy physics community, performance, and resilience to back-reaction \cite{Cacciari:2008gn}.
%{\color{blue} what exactly do you mean by
%back-reaction?}. {\color {red} this is standard %terminology to refer to the way soft particles from the UE %can modify the way the particles from the hard scattering %are clustered together, see 5https://arxiv.org/pdf/0802.1188.pdf and the antikt paper, %fig.4 https://arxiv.org/pdf/0802.1189.pdf}   
Various constituents, underlying event subtraction and corrections for the 
jet energy resolution and detector effects have been used in heavy-ion measurements.

Jet measurements in heavy ion collisions have used different constituents
for the jets.  One approach is
to only use charged particles reconstructed in the tracker as jet constituents. 
The advantage to this is that there is a clear connection between the particles
which make up the jet and the measured constituents.  Another advantage is the 
excellent pointing and angular resolution of tracks, which is
relevant for substructure measurements. The two main downsides
are that neutral particles (one third of the jet on average)
are completely excluded from track-based jets and 
that tracking generally becomes difficult when the track density is large (as
in the core of a jet) and the particles are at high-\pT\ where the track momentum 
resolution increases. Given the potential benefits of track-based measurements in terms of precision (see for instance ATLAS jet substructure measurements in Ref.~\cite{ATLAS:2019mgf}), theoretical tools such as track functions~\cite{Chang:2013rca} are being developed to analytically calculate track-based observables.

Another technique is to use purely calorimetric information, utilizing both
hadronic and electromagnetic calorimetry.  In this case,
a much more complete picture of the jet is formed (only muons and neutrinos
which carry, on average, a very small fraction of the jet energy are excluded).  
Additionally, calorimeter measurements improve with increasing energy (up 
to the point at which energy leakage becomes significant).  One issue with this is
that the calorimeter response can depend on the fragmentation pattern of
the jet (e.g. Ref.~\cite{ATLAS:2014hvo}).
Particle flow jets, first used in
ALEPH~\cite{ALEPH:1994ayc}, are commonly used in high-energy physics 
and are increasingly used in 
heavy-ion collisions as well. For instance, CMS uses particle flow reconstruction for jet physics both in \pp\ and \pbpb\ \cite{CMS:2017yfk}.

Particle flow is an optimized combination
of calorimeter and tracking information that is used to try to make the 
jet constituents closer to the actual particle constituents.  
This combines the advantage of track-based jets that the jet constituents
are directly related to the jet particles and the advantage of calorimeter
based jets that the full jet energy is measured. Additionally, particle flow facilitates pileup mitigation.  

Measurements in heavy ion collision need to deal with the large level of uncorrelated
underlying event background that modifies the jet $p_{T}$ and the jet internal structure. On
average, the underlying event shifts the jet $p_{T}$ proportionally to the jet area. The underlying event 
fluctuations increase the  jet energy resolutions more strongly for increasing $R$. 

The correction procedure is in general characterized by:
\begin{itemize}
\item An event-by-event correction of pedestal background that affects the jet $p_{T}$ and its substructure

\item A suppression of combinatorial (fake) jets, which are the jets that are reconstructed
by the algorithm but are not correlated to a  hard scattering. Similarly, combinatorial contributions to the jet substructure are suppressed.  

\item The unfolding of detector effects and residual background fluctuations
\end{itemize}

The first step requires an estimate of the average uncorrelated background per unit area, $\rho(\eta,\phi)$. A common procedure is the area-based method \cite{Cacciari:2007fd}.
An extension of the area-based method to correct jet shapes or any IRC-safe substructure
observable for the average background is also applied by the different experiments~\cite{Soyez:2012hv}. 
ATLAS has used an iterative determination of $\rho(\eta,\phi)$ based on the 
region of the detector which doesn't have candidate jets~\cite{ATLAS:2015twa}.
Other approaches, that consider a particle-by-particle subtraction are also used, see for instance \cite{Berta:2014eza,Berta:2019hnj}. 

The impact of combinatorial jets is mitigated when considering high-$p_{T}$ jets and/or small jet radius, $R$.
For low-$p_{T}$ jets and/or large $R$ different techniques have been applied like a data-driven subtraction procedure based on semi-inclusive coincidences (recoil)~\cite{Adam:2015doa}, requirements on the 
jet structure~\cite{ATLAS:2012tjt}, event mixing~\cite{STAR:2017hhs} or
machine-learning (ML)~\cite{Haake:2018hqn}. 

When measuring jet substructure, the problem of combinatorial subjet prongs emerges. When selecting two subjet prongs, for instance via the SoftDrop (SD) grooming procedure \cite{Larkoski:2014wba}, the purity of the measured prongs is not unity~\cite{Mulligan:2020tim}, and it decreases with lowering $z_{cut}$, grooming cuts or increasing jet $R$. 
To assess this problem, the strategy in recent measurements was to consider small-$R$ jets and high $z_{cut}$ grooming cuts. There is room to improve this substantially in
the near future.

The last step is the unfolding. The Bayesian~\cite{DAgostini:1994fjx}, SVD~\cite{Hocker:1995kb} or TUnfold~\cite{Schmitt:2012kp} algorithms are common unfolding tools used by the different collaborations to correct detector and residual background fluctuations in one or several dimensions. New tools based on ML are currently used to correct for detector effects in the context of high energy physics~\cite{Andreassen:2019cjw} and are yet to be explored in heavy-ion collisions.  
In some cases the measurements are not unfolded to the 
particle level and theoretical calculations
are smeared to match the data.  However, this approach prevents  direct comparisons
between results of different measurements.

\subsection{Jet tools}
\label{subsect:jettools}
In recent years, the application of theoretical and experimental jet tools 
developed by the high-energy physics (HEP) community to heavy-ion collisions has opened 
new opportunities, particularly in the area of jet substructure. In HEP, jet substructure has a broad set of applications, from tagging massive boosted particles or tuning MC generators, to testing the standard model or enhancing sensitivity to new physics \cite{Marzani:2019hun}. 
Jet substructure can be studied using the clustering history. The jet constituents are typically reclustered with the Cambridge/Aachen (CA) algorithm \cite{Dokshitzer:1997in,Ellis:1993tq}, which combines pairs of constituents/subjet prongs 
with the smallest angular separation first, leading to an angular-ordered jet tree. Then the clustering history can be undone. Each step of the declustering yields two subjets $S1$ and $S2$ with transverse momenta $p_{T,1}$, $p_{T,2}$ separated by a distance $\Delta R_{12}=\sqrt{\Delta y_{12}^{2}+\Delta\phi_{12}^{2}}$ and with $p_{T,1} > p_{T,2}$. 
The process can be iterated always unclustering the leading prong $S1$ until two subjet prongs are found that satisfy a given kinematic condition--this is what grooming algorithms do as we will describe. Or the process can be iterated till the given $S1$ cannot be unclustered (when it is a single-particle prong), to study the kinematics of all the jet prongs and build what is called the primary Lund Jet Plane \cite{Dreyer:2018nbf}. 

The massDrop/SoftDrop groomer \cite{Butterworth:2008iy,Larkoski:2014wba} stops the declustering when the following conditions are met:
\begin{equation}
z_{12}=min(p_{T,1},p_{T,2})/(p_{T,1}+p_{T,2})>z_{cut}(\Delta R_{12}/R)^{\beta}
\end{equation} where $\beta$ and $z_{cut}$ control the grooming and are choices in a particular analysis.
In essence, the algorithm removes large-angle and soft branches until a sufficiently hard splitting is found. The resulting groomed jet is less affected by non-perturbative effects like the underlying event or pileup. Another recently developed grooming algorithm is the dynamical groomer \cite{Mehtar-Tani:2019rrk}.

Alternatively, instead of using the clustering history to select hard substructure, jet
trimming~\cite{Krohn:2009th} 
consists of reclustering the jet constituents with a resolution parameter smaller than the original jet $R$ to keep only the subjets that satisfy $p_{T,sub}>z_{cut}p_{T,jet}$. Those  subjets which 
kept are merged to form the trimmed jet. 

In heavy-ion collisions, the possibility to select hard components of the parton shower to study microscopic properties of the QGP has generated a lot of theoretical and experimental interest and new synergies between the HEP and HI communities.

\subsection{Jet Observables}

In order to capture the physics of jet quenching, three different classes of observables are studied:
\begin{itemize}
\item inclusive jet suppression

\item inter-jet correlations via hadron-jet, boson-jet or di-jet coincidence measurements. 
\item intra-jet distributions, via jet shapes and jet substructure
\end{itemize}
These classes of observables are sensitive to different aspects of jet-medium interactions and must be dynamically correlated.  Here we define 
several of the common quantities used in this field.

\paragraph{Inclusive jet suppression}
The first class of observables are simply yield measurements of single jets (or hadrons).
These yield are compared to expectations from scaling cross sections in \pp\ collisions $\frac{d^2 \sigma_{jet}}{d\pT dy}$ by 
the nuclear thickness function, \TAA.  This is quantified by the
nuclear modification factor
\begin{equation}
\RAA \equiv \frac{\frac{1}{N_{evt}}\frac{d^2 N_{jet}}{d\pT dy}}{\TAA \frac{d^2 \sigma_{jet}}{d\pT dy}}.
\label{eq:RAA}
\end{equation}
For objects which lose energy traversing the QGP, \RAA\ is less
than unity but the value of \RAA\ depends on the amount 
of energy lost by the jet and the shape of the underlying \pT\ spectrum
of the objects of interest.
For a fixed energy loss,
a spectra with a steeper \pT\ dependence will have a smaller \RAA\ value.  
In the absence of a \pp\ reference dataset, \Rcp, the central to peripheral 
collision yield ratio, has been measured:
\begin{equation}
\Rcp \equiv \frac{\TAA_{per}}{\TAA_{cent}}\frac{\frac{1}{N_{evt,per}}\frac{d^2 N_{jet,cent}}{d\pT dy}}{\frac{1}{N_{evt,cent}}\frac{d^2 N_{jet,per}}{d\pT dy}},
\label{eq:RCP}
\end{equation}
where $\TAA_{cent}$ and $\TAA_{per}$ are the nuclear thickness function for
the central and peripheral events, respectively.  In all cases
\TAA\ is calculated via the Glauber model~\cite{Miller:2007ri}.

In addition to the absolute rate of jet production, the the azimuthal variation of the jet yield relative to the event plane angles
can be measured.
This is quantified via the 
vn coefficient defined as:~\cite{Heinz:2013th}:
\begin{equation}
\frac{dN}{d\phi} \propto 1 + 2\sum\limits_{n=1}^{n} \vn \cos\left( n \left(\phi - \Psi_n \right) \right)
\end{equation}
Since the event planes are understood to be driven by the geometry
of the overlap of the two nuclei in the collision, jet-\vn\ measurements
are sensitive to the  dependence of energy loss on the length of the QGP seen by the jet.

\paragraph{Inter-jet correlations}
This class of measurements study the momentum balance and the azimuthal correlation between a
jet and another object.  These observables exploit the fact that high-\pT\
objects must be produced in momentum conserving processes and that these are dominantly
2~$\to$~2 scatterings.
A common example is where one object is a high-\pT\ photon
and one is a jet.  The photon doesn't interact strongly with the QGP and thus retains
its original momentum.  The jet is produced opposite in azimuth and its momentum and direction
can be changed by interacting with the QGP.  The momentum balance is quantified
via: 
\begin{equation}
\xJgamma\ \equiv \frac{\ptjet}{\ptgamma}.
\end{equation}
Because the photon momentum is unaltered, differences in this quantity between
heavy-ion collisions and \pp\ collisions are attributable to energy loss.

The same technique can be used when both of the objects are jets.  In that
case the quantity:
\begin{equation}
\xJ \equiv \frac{p_{\mathrm{T}}^{\mathrm{jet,subleading}}}{p_{\mathrm{T}}^{\mathrm{jet,leading}}}
\end{equation}
is measured.  Here the labels \textit{leading} and \textit{subleading} denote the
highest and second highest \pT\ jets in the event, respectively.  The interpretation
isn't as simple as in the photon-jet case because the leading jet can also lose energy,
however this observable is sensitive to the difference in energy loss
between the leading and sub-leading jets.

\paragraph{Intra-jet observables}
The intra-jet observables measured in heavy ion collisions can be further classified into: 
\begin{itemize}
    \item Those that are built directly from the
    positions and momenta of the jet constituents. Examples of this are jet fragmentation functions, which provide information on how jet constituents carry the jet momentum, or what are called generalised angularities, a set of observables that can be constructed as moments of the angle of the constituents relative to the jet axis and the constituent energy. Examples of measured angularities are the jet mass, the jet girth, $g$, and the momentum dispersion, $p_{T,D}$.
    \item Those that are built using the jet clustering history. The re(de)clustering process introduces a hierarchy and is used to access parts of the jet tree that are well defined theoretically and that are expected to be connected to the gluon emissions in the parton shower process. Examples of such observables that will be discussed in this review are the groomed momentum balance $z_{g}$, the groomed jet radius $R_{g}$, the Les Houches multiplicity  $n_{SD}$, the $N$-subjettiness or the $k_{T}$ distance.
\end{itemize}

%Two examples belonging to the first class of observables are the fragmentation functions and the angularities. Measurements of fragmentation functions and jet shapes in general provide information on how the particles inside the jet carry the momentum. 
%In these measurements 
%ideally the measurement of the jet itself is independent of the measurement of the
%constituents.
Fragmentation functions are typically measured as a function of:
\begin{equation}
z \equiv \frac{\ptpart \cos \Dr}{\ptjet}
\end{equation}
where \ptpart\ is the transverse momentum of the particle of interest, \Dr\ is the 
angular distance between the jet axis and the particle and \ptjet\ is the 
transverse momentum of the jet.  In this way, \z\ represents the longitudinal momentum
fraction of the particle with respect to the jet. 
Other definitions of fragmentation functions, including 
as a function of \ptpart\ and 
in two dimensions, both angular and longitudinal directions,
have been considered.

The angularities can be described as a two-parameter family of
observables~\cite{Larkoski:2014pca}:
\begin{equation}
    \lambda_{\beta^{k}}=\sum_{i \in jet} z_{i}^{k} (R_{i}/R_{0})^{\beta}
    \label{equation:angularities}
\end{equation}
where the particular choice of parameters $(\beta,k)=(0,0),(0,2),(1,1),(2,1)$ correspond to jet multiplicity, $p_{T,D}$, girth and jet mass respectively.

Different observables belong to the second category that uses the clustering history will be described as discussed in the following sections.

\section{What have we learned from jet measurements at the LHC and RHIC?}

We have organised available heavy-ion jet data around  three physics
questions. First, what are the mechanisms responsible for the transport of energy from high-energy to low-energy modes within the QGP? Second, can we observe jets
probing free quarks and gluons within the QGP? 
And finally, what is the critical size for the QGP formation?
The first class of measurements constrains the mechanisms of jet-medium interactions since it comprises a vast set of observables that are differential in jet size, flavour, shape, substructure and in-medium path length. We reconstruct jets after they interacted with the QGP and we measure the properties of a specific selection of jets, hose which have survived. We aim to learn about flavour hierarchy in energy loss, about the role of the medium response, or the interplay between energy loss and color coherence. These aspects of jet quenching are interconnected and measurements attempt to isolate the effects, for instance, by separating large-angle and small-angle components or by selecting jets with a hard 2-prong substructure. 
The second class of observables comprise searches of large momentum transfer interactions in the medium as a proof of point-like scatterers within the QGP fluid.  The third class of measurements comprises searches for jet quenching signatures in small systems like \pPb\
collisions that display signatures of collective effects.

\subsection{Jet energy transport within the QGP}

In this section we have grouped experimental measurements into four categories according to the aspect of the jet-medium interaction mechanisms they constrain:
\begin{itemize}
\item How opaque is the QGP to jet propagation?
\item How does the amount of lost energy depend on path length?
\item How does jet quenching depend on the characteristics of the jets?
\item What happens to the energy lost from jets in the QGP?
\end{itemize}

\subsubsection{How opaque is the QGP to jet propagation?}

\begin{figure}
\centering
 \includegraphics[width=0.44\textwidth]{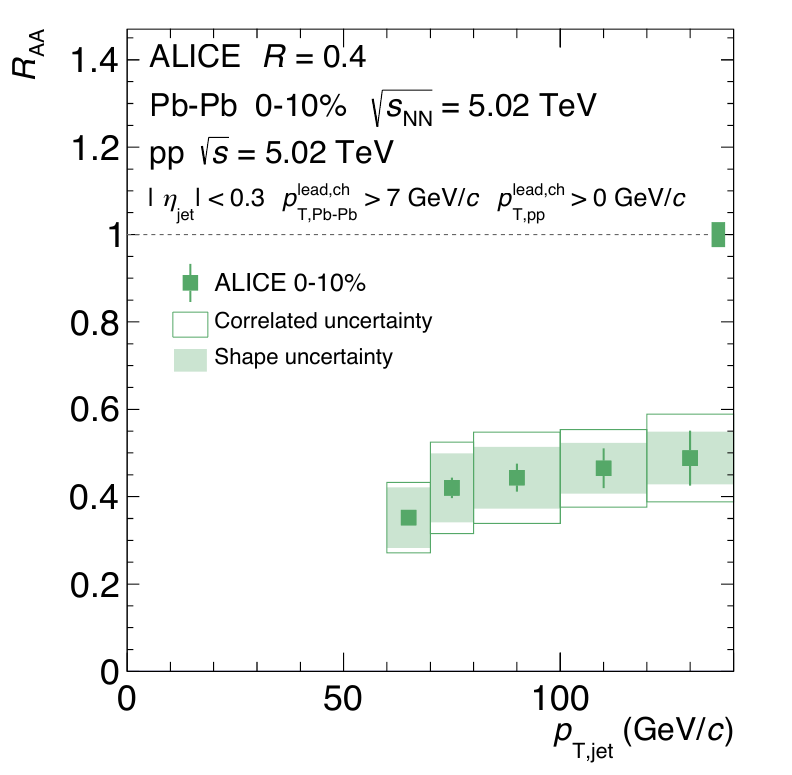}
   \includegraphics[width=0.54\textwidth]{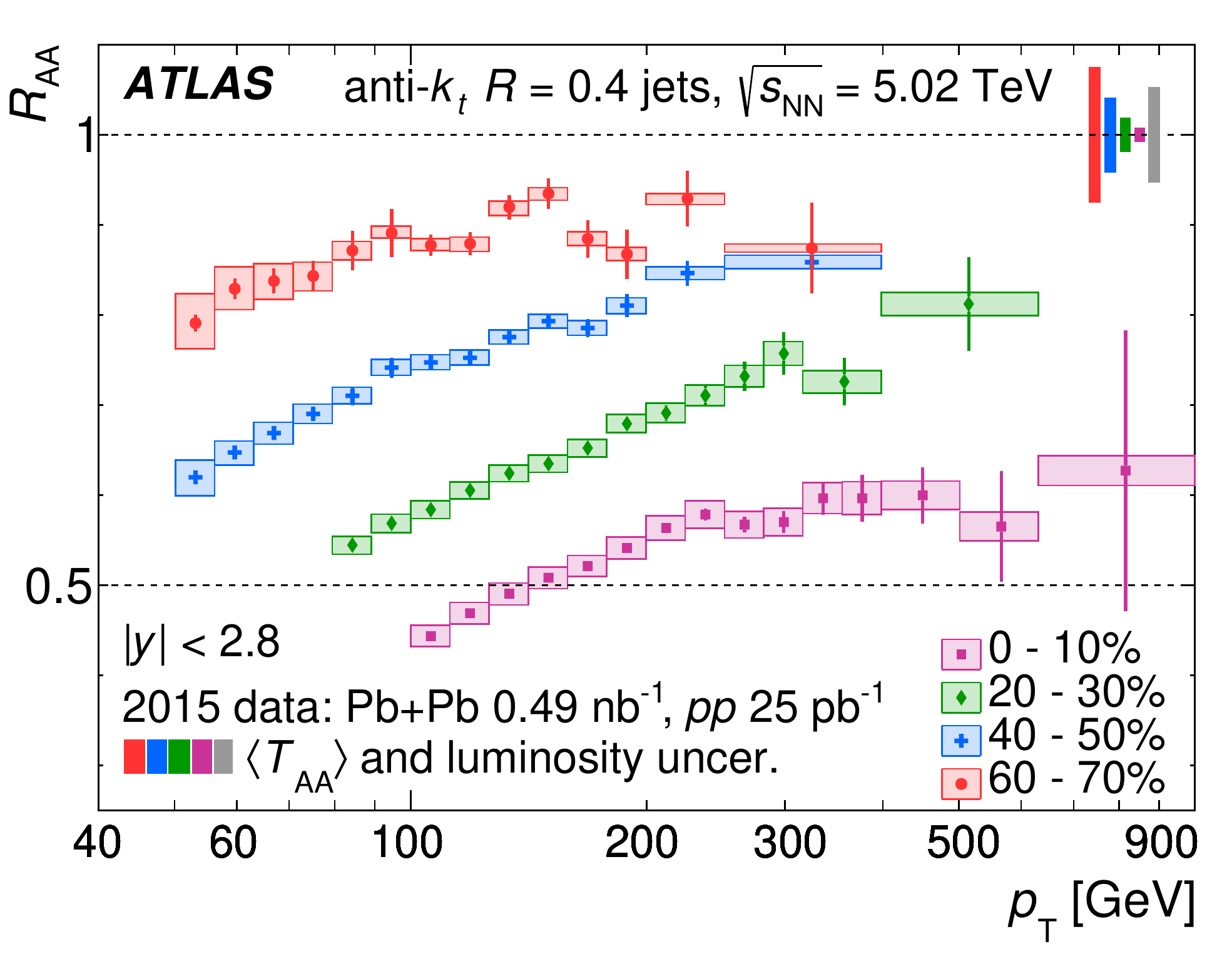}
   \caption{Jet \Raa\ as a function of \ptjet\ in \pbpb\ collisions.  Figures are from Refs.~\cite{Acharya:2019jyg} (left) and \cite{Aaboud:2018twu} (right).}
   \label{fig:incjetsRAA}
\end{figure}

\begin{figure}
\centering
\includegraphics[width=0.95\textwidth]{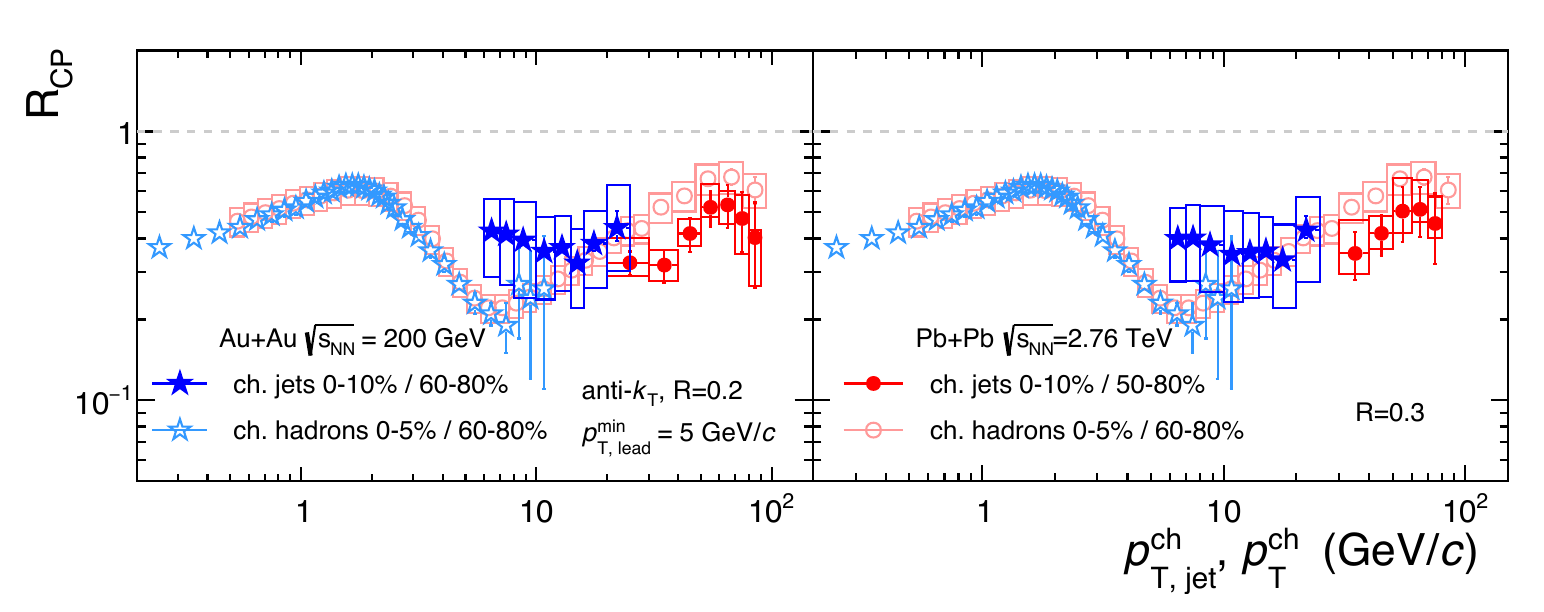}
\caption{\Rcp\ for jets and charged particles at RHIC and the LHC
(as indicated on the plot)
for \RTwo\ (left) and \RThree\ jets (right).  From Ref.~\cite{Adam:2020wen}.}
\label{fig:starrcp}
\end{figure}

Inclusive jets dominantly originate from light quarks and gluons.  The high rates of these
jets allow for measurements over a very large kinematic range. Figure~\ref{fig:incjetsRAA} 
shows recent ALICE and ATLAS results of the 
jet nuclear modification factor, \Raa, measured in 0--10\% 
central collisions from 60~GeV to 1~TeV
for \RFour\ jets~\cite{Acharya:2019jyg,Aaboud:2018twu} for 5.02~TeV \pbpb\
collisions at the LHC.  
\Raa\ remains below unity over the entire measured 
kinematic range.  
At RHIC, the jet \Rcp\ was measured~\cite{Adam:2020wen} and been found to be
consistent with values from Ref.~\cite{Abelev:2013kqa} in 2.76~TeV \pbpb\ collisions,
see Figure~\ref{fig:starrcp}. Additionally, \RAA\ has 
been measured for neutral pions~\cite{PHENIX:2012jha}.

In order to extract energy loss values from these measurements, it is necessary to have a
model.  Jet quenching models are reviewed in 
Refs.~\cite{Qin:2015srf,Blaizot:2015lma,Cao:2020wlm}.
A great deal of theoretical work has gone into the development of these
models over many years.  However, constraining models with data has been a 
challenge.
The use of Bayesian techniques to extract jet
quenching parameters is a recent but rapidly evolving field.
 This was first used in heavy ion collisions
to constrain the equation of state~\cite{Pratt:2015zsa} and is now used to extract
estimates for the QGP bulk properties including sheer and 
bulk viscosity (see recent examples in 
Refs.~\cite{Bernhard:2019bmu, JETSCAPE:2020mzn, Nijs:2020roc}).

From the LHC data, Ref.~\cite{He:2018gks} calculated that
jets have lost an average of 10--50 GeV for \ptjet\ between 100 and 900~GeV.  
These energy loss
values provide additional information based on 
the \RAA\ values, but they are 
not direct properties of the QGP.  The extraction of \qhat\ from energy
loss measurements was recently performed in Refs.~\cite{Ke:2020clc,Cao:2021keo} using
the LIDO and JETSCAPE software, respectively.
Both of these papers constrain the models to experimental data by evolving a jet
quenching calculation through a 2+1D hydrodynamic evolution (using event-averaged
initial conditions).    At high temperature, the
two \qhat\ extractions agree and constrain $\qhat / T^3$ to be approximately 1--5
over the range 300~$< T <$~500~MeV
but the result from LIDO increases
sharply to 10--15 for $T<$~300~MeV, while the result from JETSCAPE remains constant
in that same range (both of these extractions are for $p=$~100~GeV).  
The extractions use different energy loss models
and a different selection of experimental data and it is not clear
which (or both) aspect of the models leads to the low-$T$ difference in \qhat.
Both extractions include data from RHIC, but LIDO 
includes the STAR jet \Rcp\ result~\cite{Adam:2020wen} and the PHENIX \pizero\
result~\cite{Adare:2012wg} while JETSCAPE only includes the PHENIX \pizero\ result.
JETSCAPE has broken down the constraints on \qhat\ from RHIC and the
LHC data separately and shown that there is essentially no constraining power in the
RHIC data in their model due to the limited kinematic range of the measurement. 
The limited kinematic range and 
statistical precision of the available RHIC data mean that the extractions are dominated
by the LHC data at 5.02~TeV.  This should 
change with data from the sPHENIX experiment~\cite{Adare:2015kwa}.

The ability to extract \qhat\ from the data via Bayesian analysis is a substantial step
forward in jet physics in heavy ion collisions.  The current analyses represent a proof of concept of the Bayesian techniques and are improvable in several ways. On the one hand, next generation of analysis will include more differential jet observables that pose more constrain to the models than single hadron or fixed-$R$ inclusive jet suppression. On the other hand, a wider set of model calculations and approximations should be included in the analysis.  
Other aspects, like going beyond event-averaged geometry to be sensitive to geometrical fluctuations in energy loss are also to be addressed.

%Second, the geometry used in both Bayesian analyses is event averaged. This
%means that the results are  not at all sensitive to any affect from geometrical
%fluctuations in the energy loss.  

%However, 
%these analyses are so far limited in 
%several ways.  

%First, single inclusive jet or hadron data is used and this provides
%limited sensitivity to the details of the models because these observables are 
%the most inclusive jet observables in heavy-ion collisions and thus lack 
%the sensitivity to jet flavor and structure present in other observables discussed below.
%An analysis that combines multiple %types of observables sensitive to %both energy loss and transverse %momentum broadening, for instance, %could pose stronger constrains to %the theory.

%Second, the geometry used in both Bayesian analyses is event averaged. This
%means that the results are  not at all sensitive to any affect from geometrical
%fluctuations in the energy loss.  

%Thirdly, the limited kinematic range and 
%statistical precision of the available RHIC data means that the extractions are dominated
%by the LHC data at 5.02~TeV.  This should 
%change with data from the sPHENIX experiment~\cite{Adare:2015kwa}.
%This field is evolving rapidly and we expect these and other
%shortcomings to be addressed in the near future. 
%{\color{blue} this sentence seems repetitive and I would suggest removing it}

For the rest of this section, we discuss other measurements which can provide
more experimental information about the details of energy transport in the QGP.

\subsubsection{How does the amount of lost energy depend on path length?}
A fundamental question is how energy loss of jets depends on the
path length the jet travels through the QGP.
We cannot know the specific path length traveled by the jet because of:
\begin{itemize}
\item event-by-event variation of the QGP shape and size
\item the unknown position of the hard scattering process within the QGP
\item the random propagation direction of the jet within the QGP 
\end{itemize}
Because of these there can be a large variation between the path lengths encountered
even by jets produced within the same hard scattering.  This variation, 
along with the steeply falling jet cross section with transverse momentum
leads to a selection bias toward jets which have lost little energy and thus likely 
also travelled through a smaller than average amount of QGP.  This 
is called the \textit{surface bias}~\cite{Renk:2012ve}.

In addition to  path-length 
variation there can also be  fluctuations in the energy loss
process~\cite{Milhano:2015mng}.  
In order to isolate effects which are sensitive to path-length
variations, jet observables which are 
differential in the QGP geometry can be measured. 
Additionally, model calculations must incorporate 
realistic event-by-event geometry into calculations in order to make
meaningful comparisons to data.  In this section, we will discuss
the physics processes thought to govern this question and the available measurements.
We will finish with some open questions.

In the perturbative description of energy loss, the spectrum of the emitted gluons is 
expected to be $dI/d\omega \propto 1/\omega$ if the interactions with the medium are incoherent.
However the Landau–Pomeranchuk–Migdal (LPM) 
effect in the QGP~\cite{Baier:1996kr,Wang:1994fx} leads to $dI/d\omega \propto 1/\omega^{3/2}$ for
$\omega < \omega_{c}$ and this leads to a quadratic dependence of the energy loss on the
in-medium path length, $L$, $\Delta E_{loss} \propto L^{2}$.
In a nonperturbative strong coupling model $\Delta E_{loss} \propto L^{3}$ is 
expected~\cite{Chesler:2008uy}. 

\begin{figure}[ht]
\centering
\includegraphics[width=0.93\textwidth]{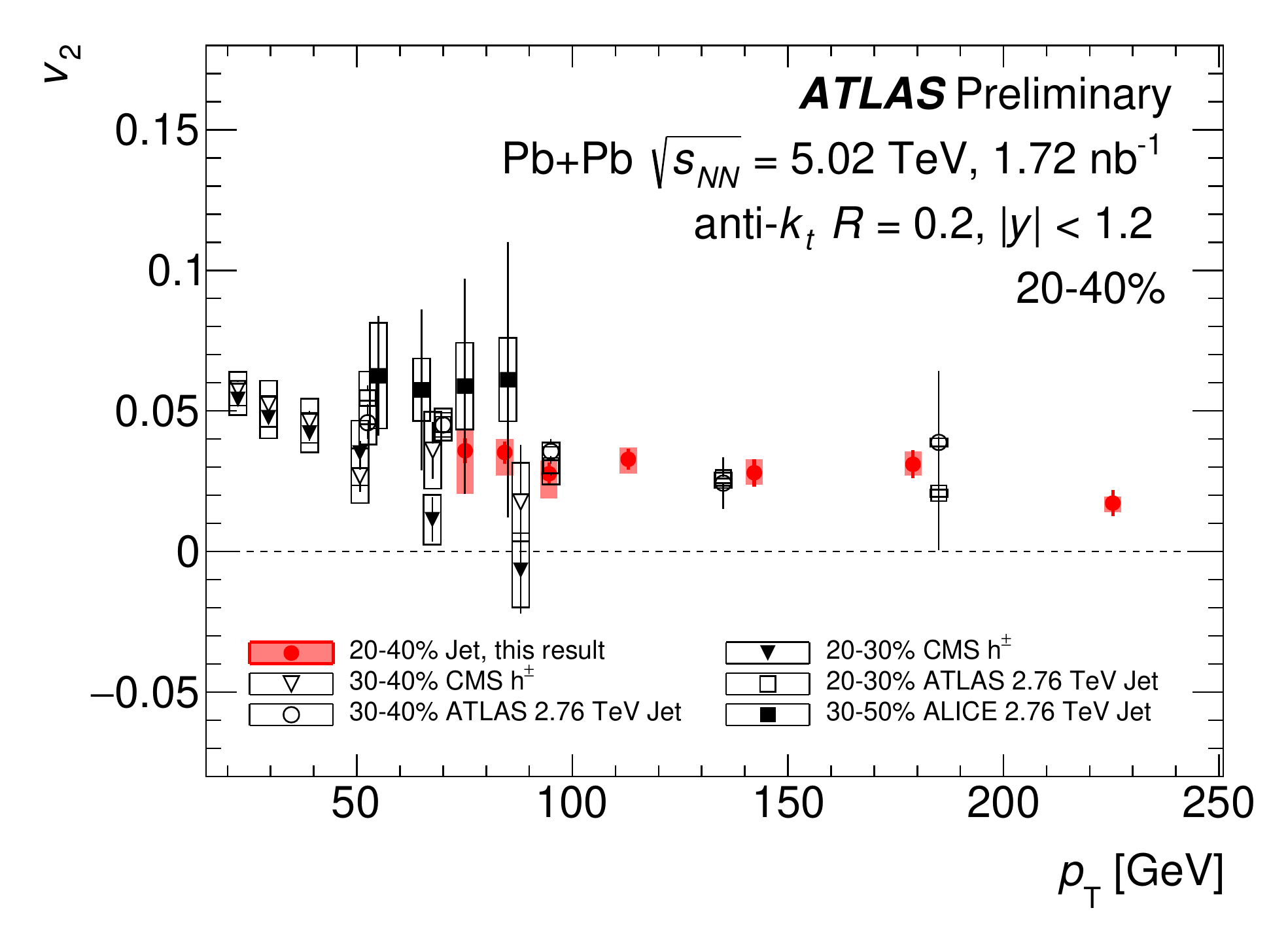}
\caption{A compilation of jet~\cite{Adam:2015mda,ATLAS:2020qxc} and 
charged-particle~\cite{Sirunyan:2017pan} \vtwo\
measurements.  Figure is from Ref.~\cite{ATLAS:2020qxc}.}
\label{fig:v2}
\end{figure}

$\Delta E_{loss}(L)$ itself is not directly measurable.  Instead, the key element in
this study has been to measure the azimuthal anisotropy, \vn, of jets and high-\pt\ particles. 
Measurements from RHIC using hadrons showed a larger $v_2$ than expected
from pQCD-based energy loss calculations~\cite{Adare:2010sp}.  This
was taken as possible evidence for strong coupling energy loss with a stronger
 dependence on $L$ than expected from pQCD.  However,
conclusions made from these measurements were shown to be limited
by the use of non-fluctuating geometry; the addition 
of geometrical fluctuations increased the value of \vtwo\ 
expected from pQCD-based theoretical calculations~\cite{Noronha-Hostler:2016eow}.
At the LHC, measurements of jet~\cite{Aad:2013sla,Adam:2015mda,ATLAS:2020qxc} and 
high-\pT\ charged particle~\cite{Sirunyan:2017pan} \vtwo\ have been performed at the LHC;
a compilation of the measurements for mid-central \pbpb\ collisions is shown in 
Figure~\ref{fig:v2}.  The \vtwo\ value varies from approximately 5\% for 20~GeV charged particles
to about 2\% for 200~GeV jets.  The \vtwo\ values as a function of centrality follow
the geometrical expectations; a smaller \vtwo\ value is seen in central collisions
than in mid-central and peripheral collisions~\cite{ATLAS:2020qxc}.  In order to further 
constrain the path-length dependence of energy loss, measurements of \vthree\ and 
\vfour\ have been made for jets~\cite{ATLAS:2020qxc} and high-\pT\ charged
particles~\cite{Sirunyan:2017pan}; above 20~GeV, there is no evidence for non-zero
\vthree\ or \vfour\ in any collision system.   These higher-order
harmonics should introduce a smaller
path length difference between in-plane and out-of-plane directions than \vtwo\
and so it is important to improve the precision of these measurements
in order experimentally constrain the path length dependence of energy loss.

Interestingly, a non-zero \vtwo\ has  been measured for high-\pT\ charged
particles in \pPb\ collisions~\cite{Aad:2019ajj}.  The measured \vtwo\ is approximately
2\% for 20--50~GeV particles in central \pPb\ collisions.  In contrast to \pbpb\ collisions,
the \vtwo\ in \pPb\ collisions is not accompanied by a large energy loss; in \pPb\
collisions \RpPb\ is consistent with unity~\cite{Adam:2015hoa,Khachatryan:2016odn}.  
If this \vtwo\ arises from path-length
dependent energy loss, the absolute size of the energy loss
would have to be sufficiently small to accommodate the
\RpPb\ results.  Thus far, there is no understanding of
whether the observed \vtwo\ can be attributed to energy loss or
if some other source is required to explain the data.  
If the \pPb\ \vtwo\ is due to some other mechanism 
than path-length-dependent energy loss, then the impact to the commonly accepted understanding of these measurements in heavy-ion
collisions needs to be assessed.

\begin{figure}[ht]
\centering
\includegraphics[width=0.50\textwidth]{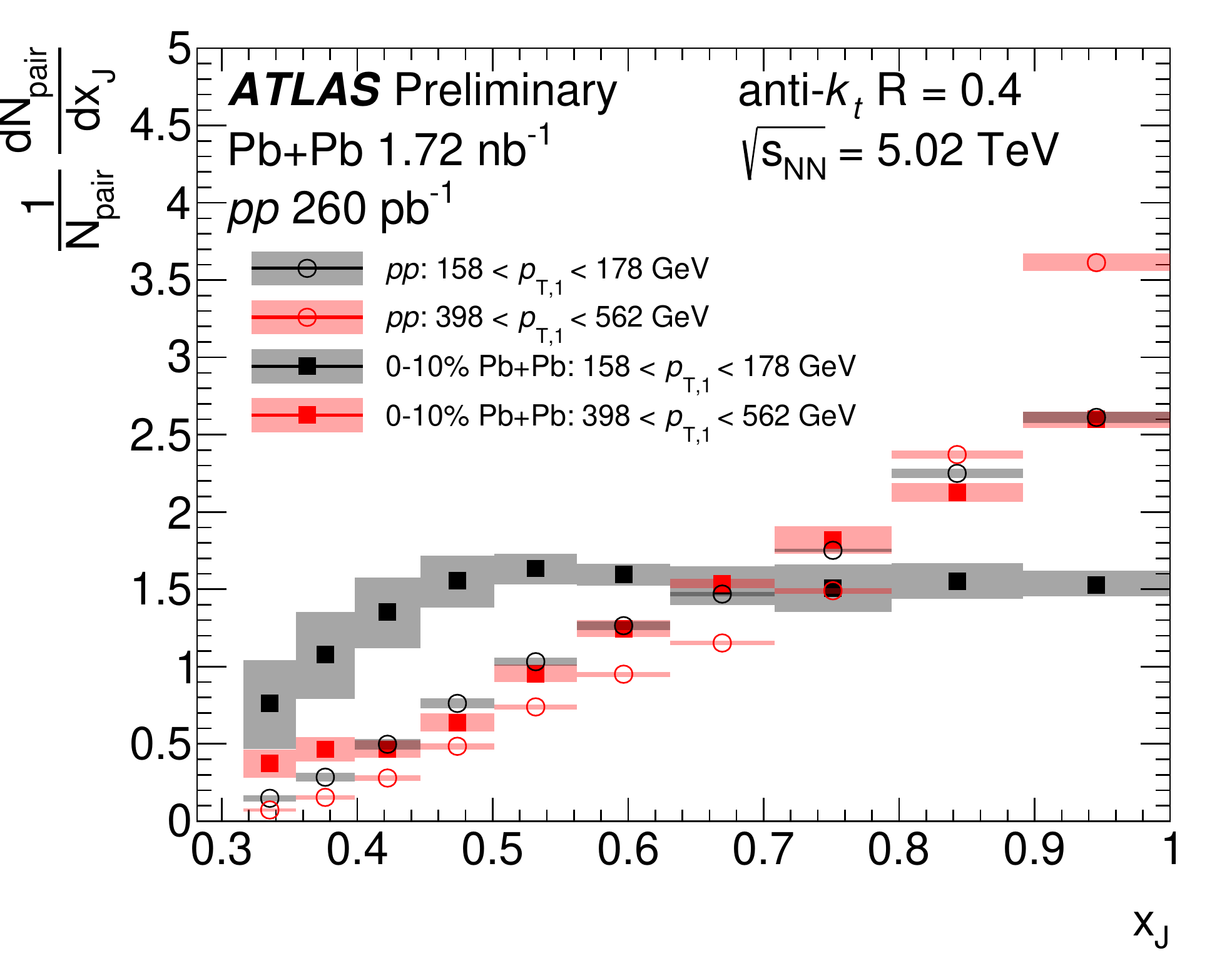}
\includegraphics[width=0.45\textwidth]{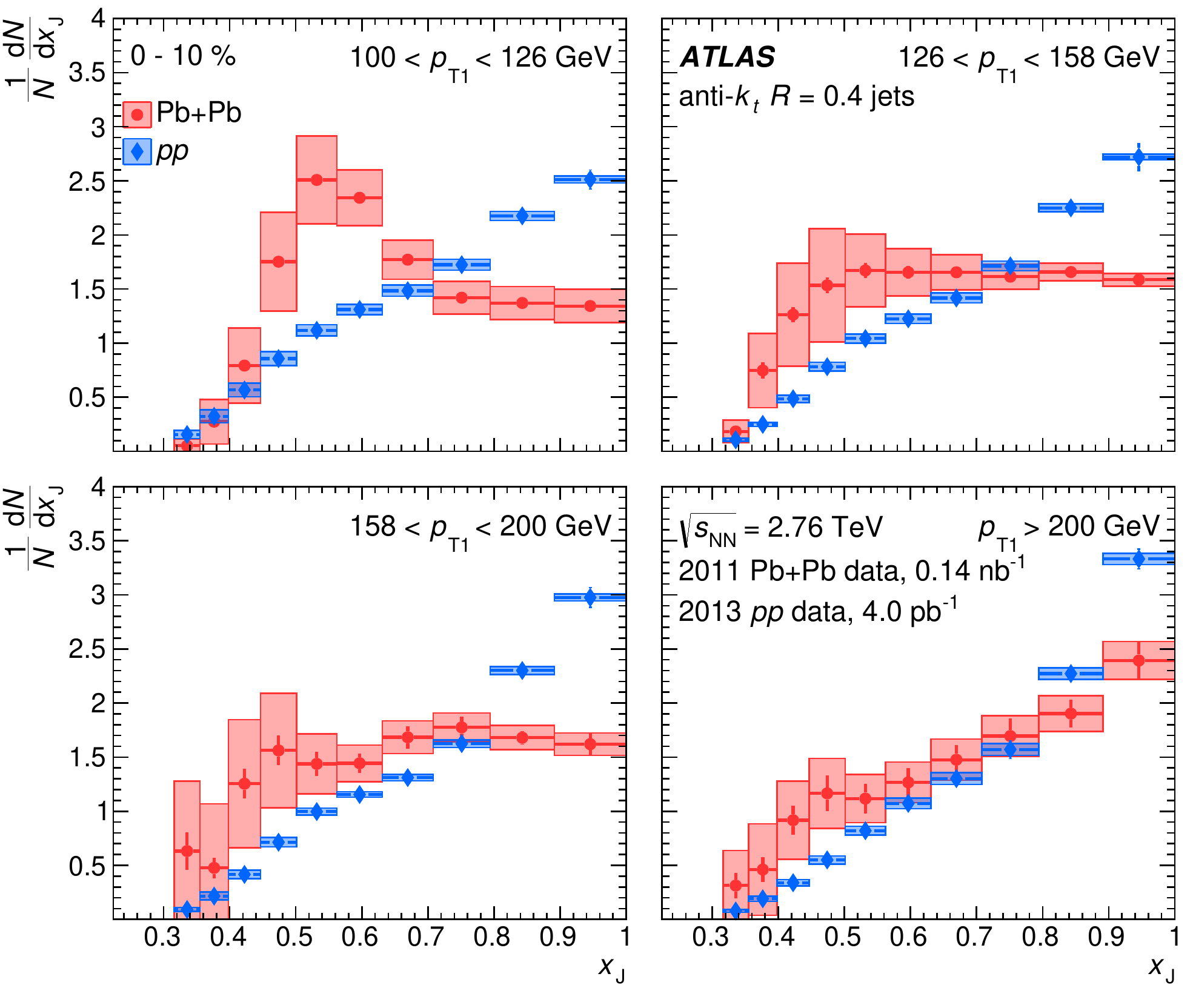}
\caption{Dijet momentum imbalance, \xJ, at 5.02~TeV (left) and 2.76~TeV (right)
for 0--10\% central \pbpb\ collisions and \pp\ collisions. Figures are from Ref.~\cite{ATLAS:2020jyp} (left) and   Ref.~\cite{Aaboud:2017eww}  (right).}
\label{fig:xJ_ATLAS}
\end{figure}

Dijet measurements provide a different sensitivity to the path-length
dependence of energy loss through geometry 
than single-jet 
measurements.  
The first LHC results showed a significant depletion of balanced dijets
in \pbpb\ collisions~\cite{Aad:2010bu}.
The qualitative explanation for this is that one jet loses more energy than
the other, either through an asymmetry in the path length or through fluctuations in 
the energy loss.  
Current measurements show the same decrease in the fraction of balanced
jet pairs in central \pbpb\ collisions compared to \pp\ collisions, 
up to leading jets of at least 400~GeV~\cite{ATLAS:2020jyp}
(see Figure~\ref{fig:xJ_ATLAS}).
For leading jet \pt of 158--178~GeV the \xJ\ distribution in central
\pbpb\ collisions is consistent with no \xJ\ dependence over the range of 0.5~$< \xJ <$~1.0.
This is a very broad distribution and suggests that there is a very wide
variation in the magnitude of energy loss experienced by the subleading jet in 
these collisions.
At 2.76~TeV, the first unfolded dijet
measurements in \pbpb\ collisions also
showed the imbalanced pairs expected from energy loss, but also an apparent peak  in
the \xJ\ distribution   at approximately 0.5~\cite{Aaboud:2017eww}.  Figure~\ref{fig:xJ_ATLAS}
shows the \xJ\ distributions (after the unfolding) in central \pbpb\ collisions
for jets from 100~GeV to over 200~GeV.  The peak structure is clear for the lowest
\pT\ jets and becomes insignificant for $\pT >$~126~GeV.  The origin of this structure
is not known. New measurements at 5.02~TeV have been unable to reach as
low in jet \pt to confirm this structure~\cite{ATLAS:2020jyp}.

Both the dijet imbalance and the jet azimuthal anisotropies 
should be especially sensitive to the effects of fluctuations in the initial geometry and fluctuations in the energy loss process.  
Due to this it is important to simultaneously experimentally constrain these quantities
and compare them with theoretical calculations.

\subsubsection{How does jet quenching depend on the characteristics of the jets?}

%Since the developing jet itself is what interacts with the QGP, it
%is natural to ask how the details of the individual jets matter.
%This can potentially provide information
%about the microscopic details of the jet-QGP interactions.
In the previous sections, jets were discussed as monolithic objects.  
Here, we discuss measurements of the jet properties that were done in order to probe the medium-modifications of the internal jet radiation pattern. Such modifications can provide information on the microscopic details of the jet-QGP interactions. 

The jet radiation pattern is explored via measurements of the jet shapes including fragmentation functions. 
We also discuss varying the \textit{partonic flavor} of jets between quarks, gluons
and heavy quarks, to test the flavour and mass dependence of jet-medium interactions. Finally, we discuss measurements of the hard jet substructure which aim at probing the building blocks of the parton shower in medium. 

%Both of these kinds of measurements
%provide information on how energy loss varies based on the characteristics
%of the reconstructed jet.  
%Finally, we look
%at observables sensitive to the shower development inside the QGP;
%these can provide insight into the partonic interactions as the shower
%develops.

{\bf Does jet quenching depend on the jet shape or are harder narrower jets quenched less?}

%After looking at the rates and correlations of jets, it's natural to ask how the particles within
%the jet are carrying the jet energy. 
Differences in the parton shower evolution
are expected to lead to different energy loss, so it is reasonable to ask if
the structure of the jets which survive is modified from jets in \pp\ collisions.
This question is intrinsically related to the quark/gluon differences discussed in the next subsection
because gluon jets on average have a broader and softer fragmentation than
quark jets.
% however, even within quark (or gluon) initiated jets there are large
%fluctuations in the parton shower structure.

\begin{figure}
\centering
\includegraphics[width=0.55\textwidth]{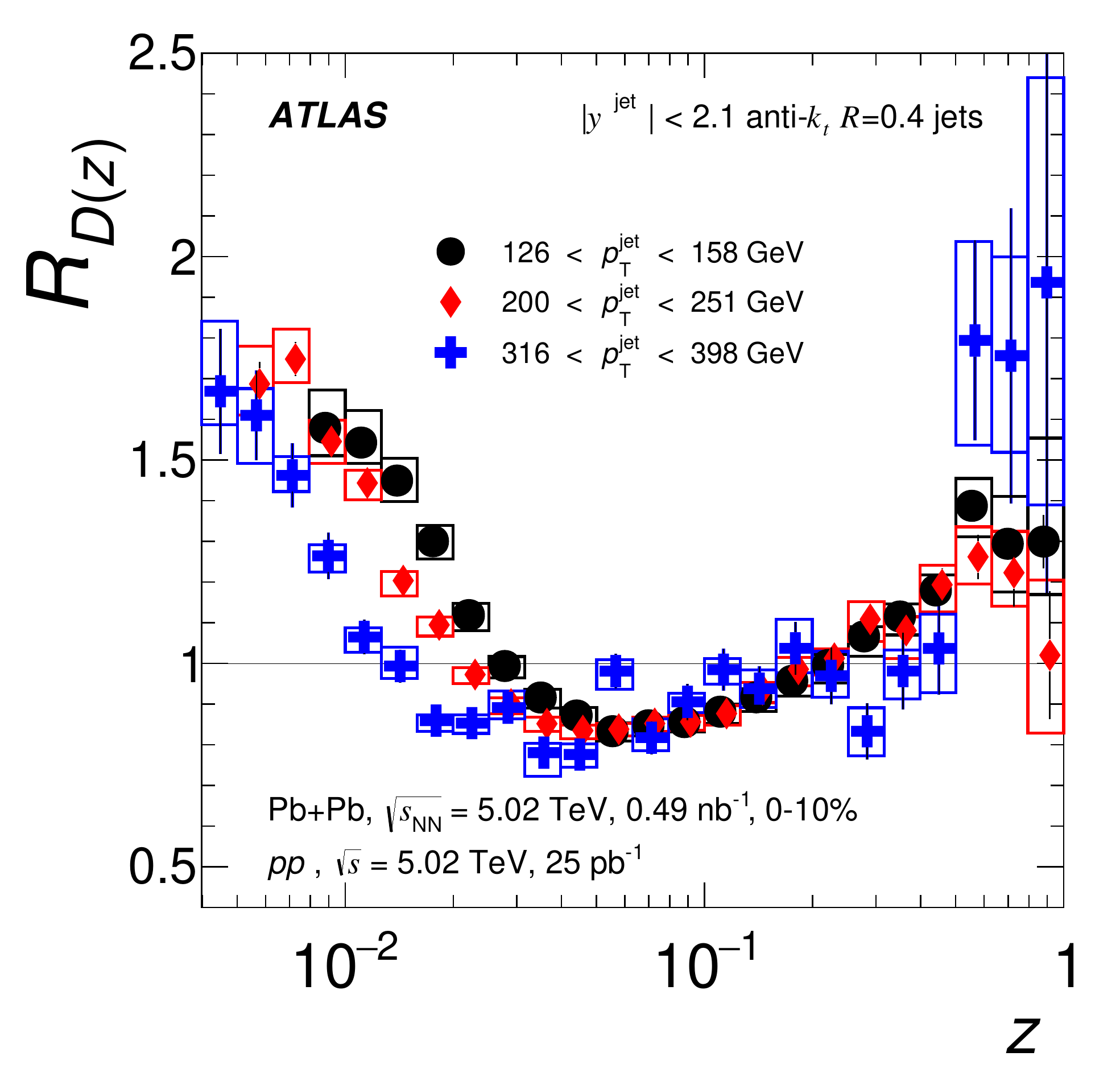}
\caption{Ratios of the fragmentation functions in 
central \pbpb\ collisions to those in \pp\ collisions
for three different \ptjet\ selections as a function of \z.  Figure is from Ref.~\cite{Aaboud:2018hpb}.}
\label{fig:FF_PbPb_z}
\end{figure}

The most comprehensive measurement
of jet fragmentation in heavy-ion collisions is in Ref.~\cite{Aaboud:2018hpb}.
Figure~\ref{fig:FF_PbPb_z} shows the fragmentation functions in central \pbpb\ events
for \RFour\ jets divided by the same quantity in \pp\ collisions for three jet
\pt\ selections.  The momentum fraction of the jet carried by the charged particle,
\z, is determined with respect to the \textit{observed} jet energy (as opposed
to the original, pre-quenching, parton energy).
At high-\z\ the ratios of the fragmentation functions are consistent
for all three jet \pt\ selections and there is an excess of high-\z\ particles in \pbpb\
collisions. This excess can be explained as the result of a selection bias: the measured jets with high-$z$ hadrons are jets with a harder fragmentation that have 
been quenched less.
Since quark jets have harder fragmentation on average than gluon jets, this
could also be understood as
evidence for a stronger energy loss for gluon jets than quark 
jets~\cite{Spousta:2015fca} resulting on an enhanced quark fraction at given \ptjet.

\begin{figure}[ht]
 \centering
 \includegraphics[width=0.85\textwidth]{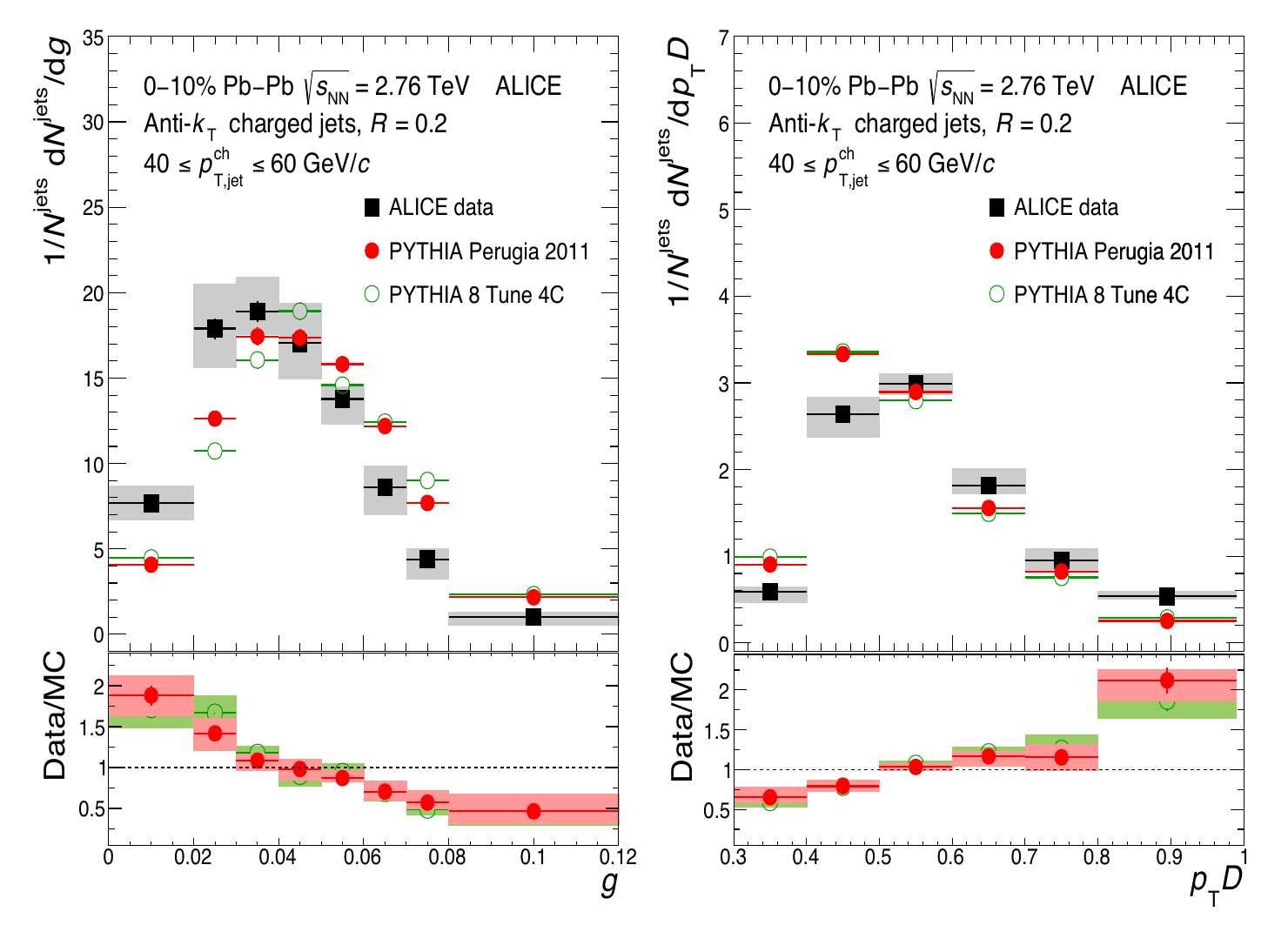}
 \caption{Jet girth and momentum dispersion in central Pb-Pb collisions compared to a vacuum simulation, from Ref.~\cite{Acharya:2018uvf}.}
 \label{fig:alice_shapes}
 \end{figure}

In order to look at the angular distribution of energy in jets, jet angularities and other jet shapes have been measured~\cite{Chatrchyan:2013kwa,Acharya:2018uvf}.  The measurement in 
 Ref.~\cite{Chatrchyan:2013kwa} measures
 the distribution of calorimeter energy inside the jets.  
 Ref.~\cite{Acharya:2018uvf} is based on unfolded 
 \RTwo\ track-based jets and includes $\ptpart >$~0.15 GeV and measures
 both the angularity, $g$ and the momentum dispersion $p_{T}D$.  These observables correspond to $\lambda_{0,2}$ and $\lambda_{1,1}$ in Equation \ref{equation:angularities} respectively.  
  The small cone size of these jets emphasizes the core and 
 minimizes the effect of any medium response.  The distributions of
 these quantities in \pbpb\ collisions are shown in Figure~\ref{fig:alice_shapes}  and indicate that the measured quenched jets are narrower and have a harder fragmentation than the \pythia~\cite{Sjostrand:2014zea}
 simulation of jets in the vacuum. 
 This can, yet again,
 be interpreted as a selection bias by which broad jets with a softer fragmentation are more quenched and are filtered out from the selected reconstructed jet $p_{\rm{T}}$ bin. 
 Interestingly, like in the fragmentation function measurements from ATLAS above, a harder and narrower fragmentation 
 is consistent with a more quark-like fragmentation and the results agree well with \pythia quark distributions as shown in~\cite{Acharya:2018uvf}. 
It is however worth noting that the measurement of the jet charge \cite{CMS:2020plq} does not indicate a change of quark and gluon fractions in \pbpb\ relative to \pp\ collisions.
 
{\bf Does jet quenching depend on quark flavour and mass?}

 At leading order in vacuum QCD, differences between quark and gluon fragmentation are dictated by color factors: the splitting rate is
 enhanced by the color factor and is 2.25
 times higher for gluons than for quarks, leading to broader and softer parton showers.  The
 larger splitting rate leads to an expectation of
 more interactions between gluon jets and the QGP.  
 %It is of interest to investigate this in order to have a 
 %solid expectation for novel medium-induced effects in the QGP.

Inclusive jets are mixture of quark and gluon jets.  The mixture is governed by
the parton distributions functions (PDFs) of the colliding
nucleons.  At low-$x$, gluons dominate and toward the valence region
there is a greater fraction of quarks.
An attempt to measure the quark and gluon fractions in 
 jets in \pbpb\ collisions has not found any 
significant difference from that measured in \pp\ collisions~\cite{Sirunyan:2020qvi}, 
but such measurements have substantial systematic uncertainties and model dependence. 
Other techniques have been
used to attempt to enhance the quark-jet fraction and to look at
the effect on the jet quenching.
One technique to enhance the quark-jet sample is to measure
the rapidity dependence of jet observables. At forward rapidities, 
the fraction of quark initiated jets will be enhanced because the jet
partons come from higher-$x$ partons than at smaller rapidities.
Alternatively, one can consider jets recoiling from isolated photons or Z-bosons. \pythia simulations \cite{CMS:2021vsp} indicate quark fractions nearly a factor three higher in Z-jet events than in central dijet events for $R=0.4$ jets of $p_{\rm{T}}<200$ GeV at 13 TeV.  
Lastly, heavy flavour jet tagging allows to study the effect of large 
quark mass on jet quenching.

\begin{itemize}
    \item \underline{Rapidity dependence of energy loss}
    Figure~\ref{fig:incjetsRAA_rap}
 shows the first evidence for
a rapidity dependence of \Raa~~\cite{Aaboud:2018twu}.  
There are two competing effects that could be expected.
First, the gluon jet fraction in the inclusive jet sample decreases toward increasing rapidity at
fixed jet transverse momentum, \ptjet, (see, for example Ref.~\cite{Spousta:2015fca} where the PYTHIA8 calculations show that the quark fraction almost doubles at forward rapidities $1.2 < |y| <2.1$  compared to $|y| <0 .3$ for jets of $p_{\rm{T}}=100$ GeV.) 
As quarks are
expected to lose less energy than gluons in the QGP, the 
value of \Raa\ would
be expected to increase as $|\yjet|$, and thus
the fraction of quark jets in the inclusive jet sample at a fixed \ptjet, 
increases.  Second, the \ptjet\ spectra become steeper with
increasing $|\yjet|$ (see, for example Ref.~\cite{Aaboud:2017dvo}); this would cause a reduction in the \Raa\
value for the same energy loss.  Figure~\ref{fig:incjetsRAA_rap}, \Raa\ is shown to decrease with increasing
$|\yjet|$ for jets with $\ptjet >$~300~GeV,
suggesting that the second effect dominates for these jets.

\begin{figure}
 
 \centering
   \includegraphics*[width=0.54\textwidth]{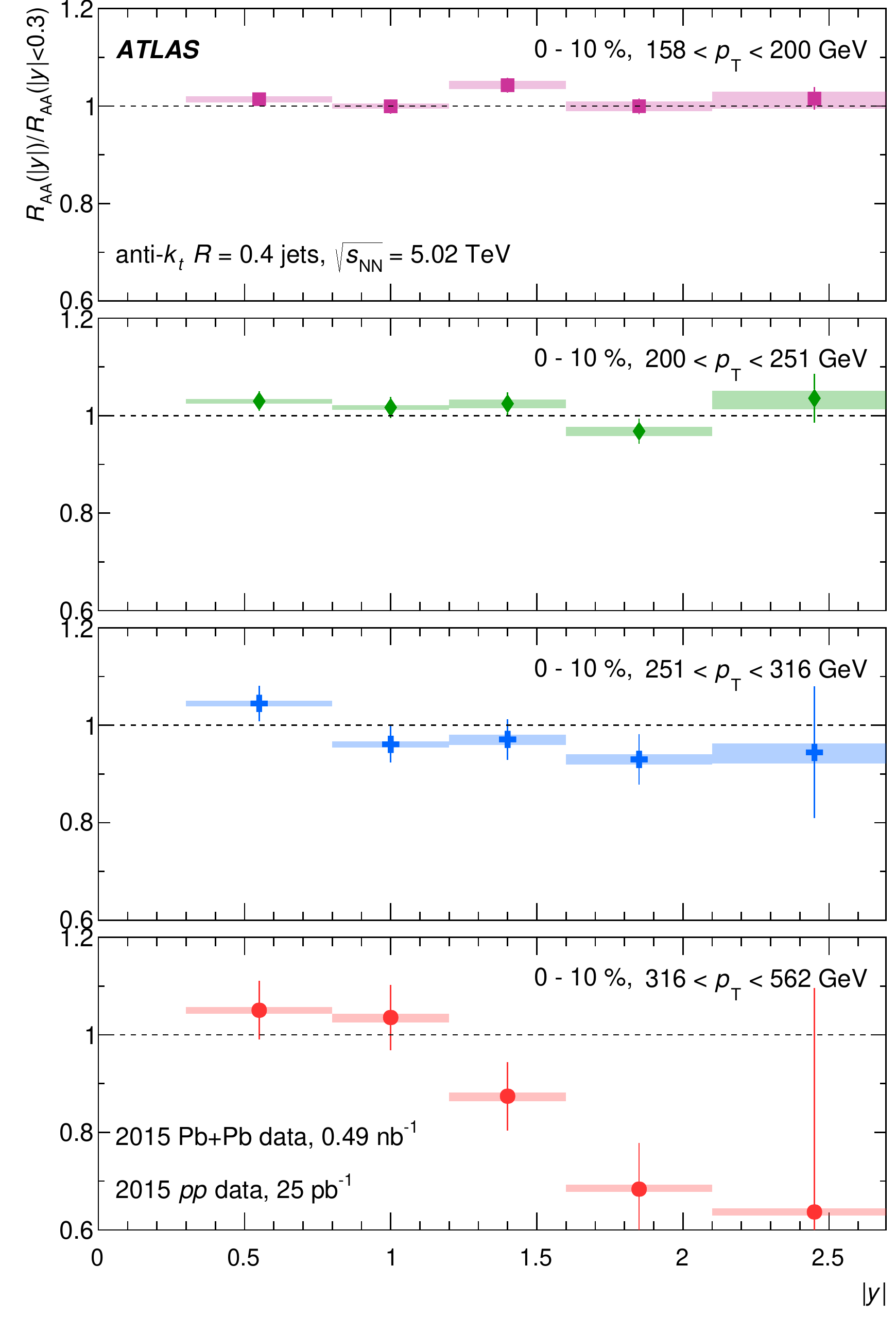}
   \caption{
      \Raa\ as a function of the jet rapidity 
      normalized by $\Raa(|y| < 0.3)$
      for four \ptjet\ selections. Figure is from Refs.~\cite{Aaboud:2018twu}.}
   \label{fig:incjetsRAA_rap}
\end{figure}

    \item \underline{Photon-tagged jet observables}
    
    Jets opposite in azimuth from a high momentum photon can also provide 
    an enhancement of quark-jets over inclusive jets because these photon-jet
    pairs are  primarily produced via $g+ q \to \gamma + q$ scattering.
    Additionally, 
the photon does not lose energy in the QGP via the strong interaction and therefore
provides information about the initial hard scattering momentum transfer.
However, these jets have a different geometrical bias than inclusive jets;
 since the photon does not lose energy, the geometrical bias toward jets
produced near the surface is removed for photon-jet measurements.

Previous measurements have shown the \pt\ of the jet relative to that of the photon is reduced
in heavy-ion 
collisions relative to \pp\ collisions~\cite{Chatrchyan:2012gt}.  
Measurements allow the study of the photon-jet
momentum balance as a function of the photon \pt~\cite{Sirunyan:2017qhf,Aaboud:2018anc}.  The ATLAS
results are unfolded and 
are shown in Figure~\ref{fig:photonjets} for 100-158~GeV photons in 0--10\% central collisions.
Going from peripheral to central collisions, the fraction of balanced photon-jet
pairs (those with $\xJgamma \approx 1$) decreases and the fraction of pairs in which the photon
has more \pT\ than the jet increases.  This is 
qualitatively as expected from jet quenching, but due to the different
geometrical bias and observable than the inclusive jets it is not possible
to say without a model if these quark-enhanced jets have lost less energy,
as would be expected.  The most probable value
of \xJgamma\ in the most central collisions is about 0.3, indicating that
many jets have lost a large fraction of their transverse momentum.  However, it is 
interesting that even in the most central collisions, there remain 
a substantial fraction of jets which are nearly
balanced--indicating that they have not lost a large amount of energy.  
Measurements with $Z$-bosons as the tag have also been done~\cite{CMS:2017eqd};
the message is similar to that of the photon-tagged measurements but the 
statistical precision is worse. 
%There results were compared to different theoretical calculations both in \pp \, and \pbpb \ collisions,that show the ability of the models to describe the baseline vacuum reference and the sensitivity of the observable to the strength of the coupling between the jet and the medium.

\begin{figure}[ht]
\centering
      \includegraphics[width=0.95\linewidth]{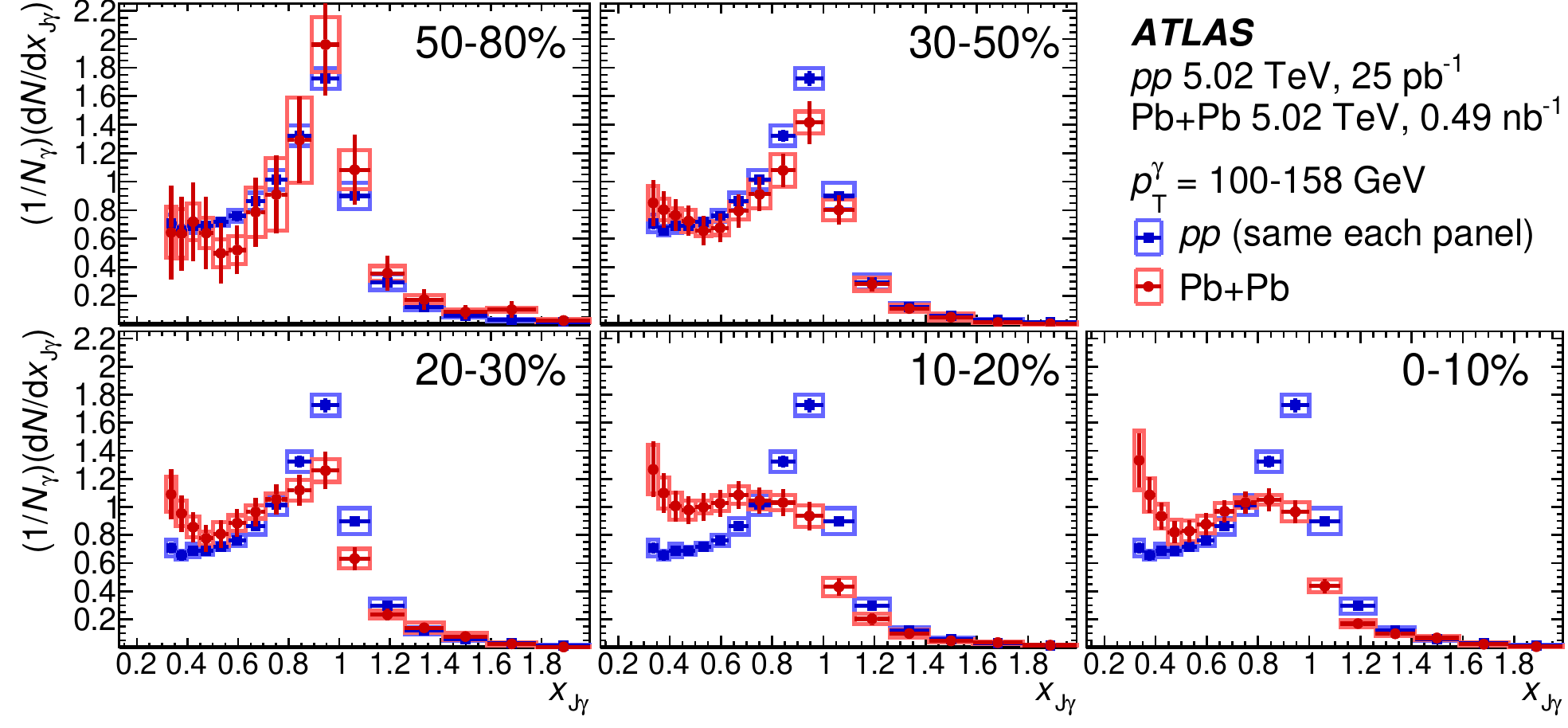}\\
      \caption{ Ratio of the jet transverse momentum to the 
      photon \pt, $x_{J\gamma}$, in \pp\ and \pbpb\ collisions~\cite{Aaboud:2018anc}. }
      \label{fig:photonjets}
\end{figure}

 First measurements of
the fragmentation of the jets opposite a photon have been performed~\cite{Sirunyan:2018qec,Aaboud:2019oac}.
As discussed above, 
these fragmentation functions differ from inclusive fragmentation functions in a few ways. 
First, the 
jets are possibly quenched more due to geometrical bias from the photon selection.
Second, the jets are at  lower \ptjet\ than the inclusive jet fragmentation functions
because the tagging with the photon limits the statistics and provides a cleaner identification
of jets at lower \ptjet\ than in the inclusive case. 
%Additionally, the \pt\
%reach of the photons is limited and that limits the \pt\ reach of the jets
%measured opposite to them.
Finally, these jets have a much higher fraction of quark jets than the inclusive jet sample
do to the leading order dominance of the $q + g \to q + \gamma$ process in these events.
Measurements of photon-hadron correlations had been made at RHIC~\cite{Adare:2012qi,Abelev:2009gu,Ge:2017irb}, 
but
only recently were measurements made of the hadrons in reconstructed jets back-to-back with
a photon in \pbpb\ collisions~\cite{Sirunyan:2018qec,Aaboud:2019oac}.
Figure~\ref{fig:photonjetFF} shows the ratio fragmentation functions
in \pbpb\ collisions compared to \pp\ collisions for both jets opposite a photon
and inclusive jets.  A stronger deviation of this ratio from unity is seen in
central collisions than in peripheral collisions for both jet selections.  Interestingly, 
when comparing the central data directly to the peripheral data, the centrality
dependence is significantly larger in the photon-tagged jets than in the inclusive jets.
It is not known if this is caused by the lower
\ptjet\ range for the photon-tagged fragmentation functions or the different
geometrical biases of the two samples, but being able to measure the fragmentation of
photon-tagged jets at the same \ptjet\ has inclusive jets would be an obvious
way to constrain the source of this difference.

\begin{figure}[ht]
\centering
      \includegraphics[width=0.95\linewidth]{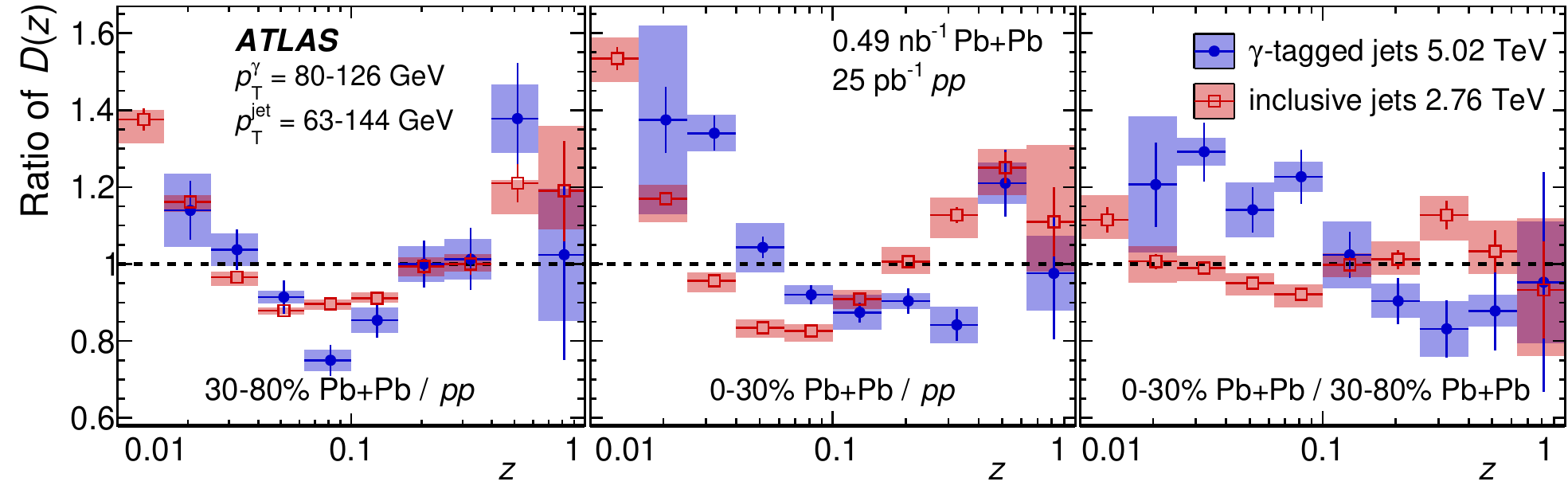}
      \caption{The ratio of the fragmentation function as a function of
      charged particle \pt\ in central \pbpb\ collisions to \pp\ collisions 
      for jets opposite a photon 
      (squares) and inclusive jet fragmentation functions~\cite{Aaboud:2017bzv} 
      (circles). Figure is from Ref.~\cite{Aaboud:2019oac}.}
      \label{fig:photonjetFF}
\end{figure}

    \item \underline{Heavy Flavour}
    
    In vacuum, besides the aforementioned differences between the radiation pattern of jets initiated by light quarks and gluons, dictated at LO by the color factors, quark mass plays a role. In QCD (an in gauge theories in general), radiation off a massive quark Q is suppressed in a cone of angle $\theta_{C}=m_{Q}/E_{Q}$. This is the 
    so-called dead cone effect \cite{Dokshitzer:1991fd} that causes heavy quarks to radiate less than light quarks. In heavy ion collisions, medium-induced radiation off heavy quarks is expected to fill the dead-cone region, but is predicted to be suppressed for high energy radiation as compared to light quarks \cite{Armesto:2003jh} resulting on a quark mass-dependence to energy loss.

The measurement of energy loss of heavy flavor jets
is very challenging.  The overall rate of these jets is very low (a few percent of the 
inclusive jet cross section) and identifying them relies on measurements sensitive to the decay of the $B$ or $D$ hadron carrying the quark
of interest inside the jet. 
CMS has made a measurement of the \bjet\ \RAA\ in 2.76 TeV
collisions~\cite{Chatrchyan:2013exa} and found consistent \RAA\ values 
between inclusive and 
\bjets.  Additionally, they measured the momentum imbalance of back-to-back \bjets\
and found them to be comparable to those measured for inclusive jets in 5.02~TeV \pbpb\ collisions~\cite{Sirunyan:2018jju}, see Figure~\ref{fig:bdijet}. Both 
measurements have sizeable uncertainties and
\bjet\ measurements will be an important part of the LHC physics program in Run 3 and beyond.

\begin{figure}[ht]
\centering
      \includegraphics[width=0.95\linewidth]{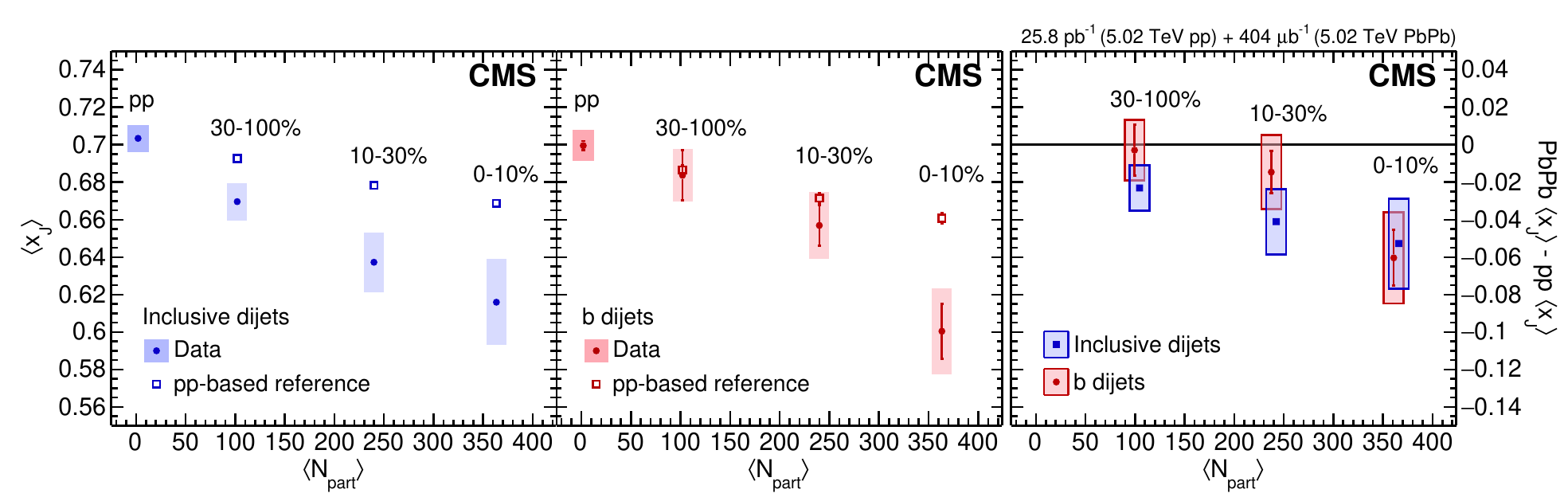}
      \caption{The momentum imbalance in inclusive and b-dijets as function of collision centrality in \pbpb \,collisions compared to \pp \,. Figure is from Ref.~\cite{Sirunyan:2018jju}}. 
      \label{fig:bdijet}
\end{figure}

The application of substructure techniques to heavy flavour jets in \pp\ collisions has recently lead to the first direct observation of the dead cone in QCD~\cite{ALICE:2021aqk}. The exploration of such techniques in heavy ion collisions is yet to happen. 
Substructure of double HF-tagged jets (discussed in Ref.~\cite{Ilten:2017rbd} in the context of disentangling heavy flavour processes in pp collisions) might allow to identify the c$\overline{\rm{c}}$ or b$\overline{\rm{b}}$ antenna without the ambiguities of the inclusive SD analysis that are subject of strong contamination of combinatorial prongs. 

%Additionally, at sufficiently high energies, HF-jet tagging becomes a way to study light-quark jet properties with high purity since mass effects are suppressed because the dead cone becomes marginal (for instance $\theta_{C} \simeq 0.04$ radians for an emission off a c-quark of 100 GeV so a c-quark jet behaves as a light-quark jet).  
%{\color{red} actually I've removed the paragraph oon HF as light quark taggers, since it is not very clear/correct. OK for you? Even for a 1 TeV c-jet, one can find 1->2 prongs with the declustering where the parent energy is low and thus register emissions that are affected by the mass}

Heavy quark measurements are expected to be substantially improved in the near
future with higher luminosity and detector upgrades 
at the LHC Runs 3 and 4~\cite{Citron:2018lsq} and the ability
to tag \bjets\ at sPHENIX at RHIC~\cite{PHENIX:2015siv}.

\end{itemize}

{\bf Does jet quenching depend on the hard substructure?}

As described in Section~\ref{subsect:jettools}, the grooming procedure stops when the SD condition is met. The corresponding $z_{12}$ and angular separation $\Delta R_{12}$ are called groomed momentum balance and groomed jet radius and are denoted by $z_{g}$ and $R_{g}$ respectively.
In vacuum, $z_{g}$ is connected to the Altarelli-Parisi splitting function and displays 
a universal behavior in $1/z$~\cite{Larkoski:2017bvj}. In Pb-Pb collisions,
the interpretation of the observable is more difficult because medium-induced radiation is expected to violate angular ordering~ \cite{Mehtar-Tani:2011hma} 
while the CA reclustering forces angular ordering on the jet constituents, among other reasons. 

Several different mechanisms can contribute to the modification of $z_{g}$
and $R_{g}$ in heavy ion collisions. If medium-induced radiation is hard enough, it can increase the number of prongs that pass the SD cut. On the other hand, jet prongs and constituents lose energy in the medium, which can 
reduce the number of subjet prongs passing the SD cut, $n_{SD}$. 
In addition, the amount of jet energy loss is dictated by color coherence: jets with a resolved substructure will lose more energy because they contain more prongs that interact with the medium incoherently.

%New formulations of the parton shower in medium \cite{Caucal:2018dla} consider a scenario where vacuum and medium-induced emissions are factorized and where the number of prongs that will interact with the medium is set by the number of vacuum prongs at the end of the vacuum shower, which happens earlier in time. 

%In addition, quantum color coherence dictates that the amount of energy lost by a jet depends on the amount of substructure prongs the medium can resolve

\begin{figure}[ht!]
\centering
      \includegraphics[width=0.8\linewidth]{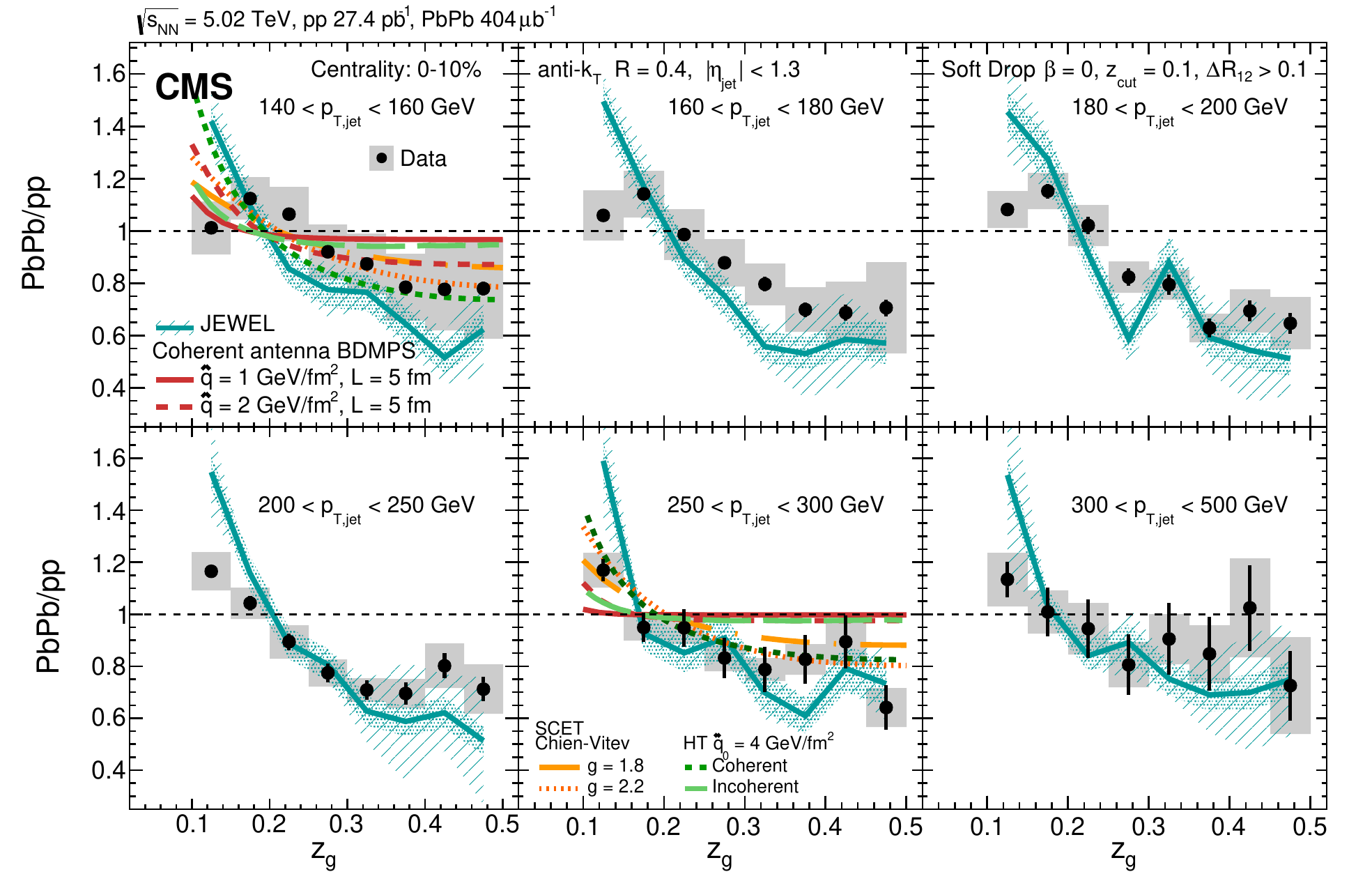}
      \includegraphics[width=0.8\linewidth]{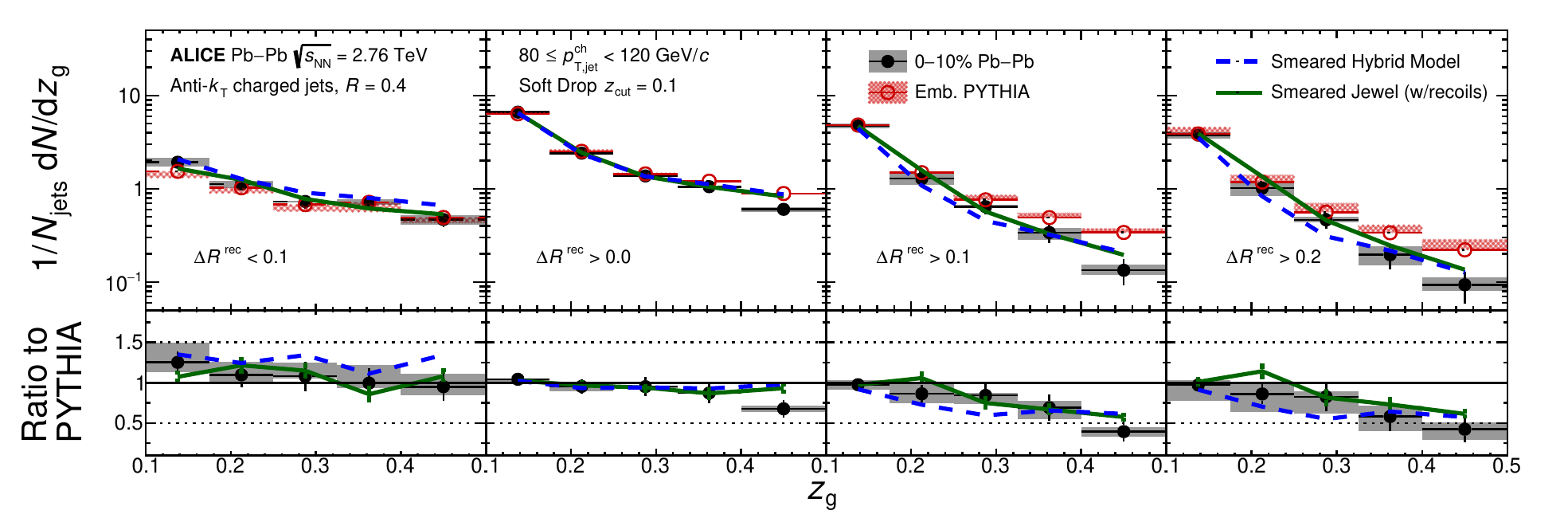}
      \caption{Upper plot: CMS self-normalized results for the momentum balance $z_{g}$ in different jet momentum bins.
      Figure is from Ref.~\cite{Sirunyan:2017bsd}. Lower plot: ALICE $z_{g}$ results for jets in a fixed momentum interval of $80<p_{T,jet}^{ch}<120$ GeV and as function of the groomed splitting angle $R_{g}$ ($\Delta R^{rec}$).
      Figure is from Ref.~\cite{Acharya:2019djg}.}
      \label{fig:zg01}
\end{figure}

%The CMS results are self-normalized and thus are only sensitive to the change of shape of the $z_{g}$ distribution. The ALICE normalization is per jet and consequently the results are also sensitive to the absolute yield. 
The $z_{g}$ distribution in heavy-ion collisions
was first measured by CMS \cite{Sirunyan:2017bsd}, then by STAR~\cite{Adam:2015doa} and ALICE~\cite{Acharya:2019djg}. 
The CMS and ALICE measurements are shown in Figure~\ref{fig:zg01}. 
The main feature of the ALICE  track-based measurement was a suppression of the $z_{g}$ distribution with increasing $R_{g}$ and a hint of an enhancement at small angles. CMS did not perform a scan on the splitting opening angle (the default value is $R_{g}>0.1$) but did examine the jet $p_{T}$ dependence of the modification. 
Both of these measurements were not fully corrected to particle-level. Rather, the \pp\ reference was modified to consider the impact of strong combinatorial background at the level of subjet prongs that dominates the low-$z_{g}$ region. ALICE also reported the measurement of the Les Houches multiplicity $n_{SD}$ which gives the number of prongs within the jet that pass the SD cut. The $n_{SD}$ distribution is shifted to lower values in PbPb relative to the vacuum calculation as expected if energy loss of the prongs reduces the number of subleading prongs passing the SD cut.

\begin{figure}[h!]
\centering
      \includegraphics[width=0.5\linewidth]{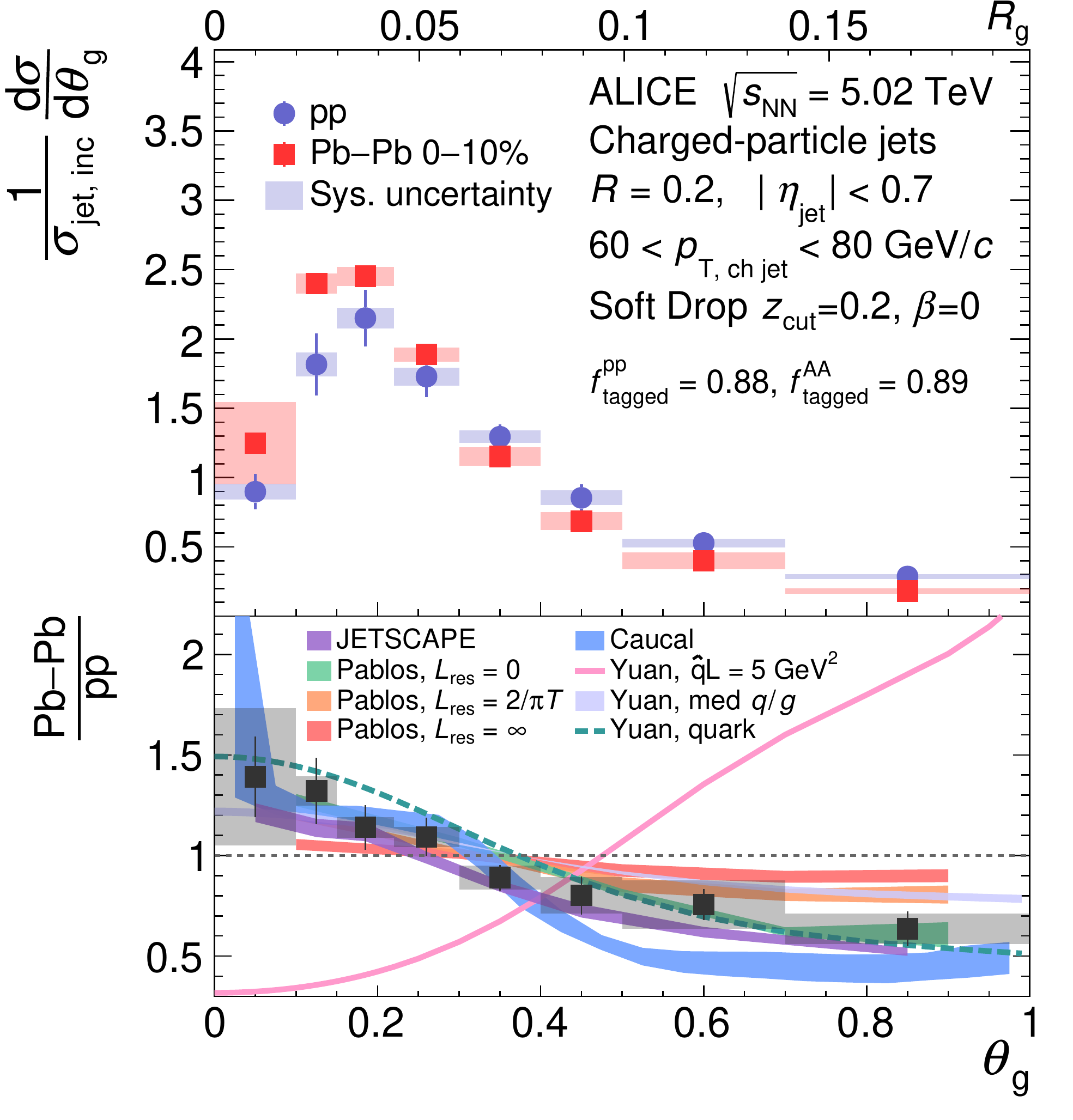}
          \caption{Normalized groomed jet radius $\theta_{g}$ in central collisions and small-R jets measured by ALICE~\cite{ALICE:2021obz}, 
          compared to the same observable measured in pp collisions and state of the art model and theory calculations. }
      \label{fig:zg02}
\end{figure}

The next generation of groomed observables by ALICE were fully corrected and, for the sake of the unfolding stability, performed with a different selection of smaller jet $R$ and tighter SD grooming cuts ($z_{cut}=0.2,z_{cut}=0.4$) \cite{ALICE:2021obz}. The results are shown in Figure \ref{fig:zg02}. A similar message is distilled: small-angle splittings are enhanced while large-angle splittings are suppressed. And the $z_{g}$, when integrating over all angles, shows no modifications.

The data were compared to a set of models, including JetMed (denoted as Caucal et al), the Hybrid model (denoted as Pablos et al) and JETSCAPE~\cite{Kauder:2018cdt}.
%In the JetMed factorized approach, 
%the number of prongs that will interact with the medium is set by the number of vacuum prongs at the end of the vacuum shower, which happens earlier in time. The Hybrid model considers two extreme cases for color coherence, $L_{res}=0$ and $L_{res}=\infty$ that correspond to the incoherent and coherent limit respectively. Non-zero $L_{res}$ means smaller effective number of in-medium emitters and yields smaller modifications of the $R_{g}$. The JETSCAPE model combines medium-modified shower at high virtualities with  Linear Boltzmann Transport model at low virtualities and describes the main features of the data. 
The narrowing of $\theta_{g}$ is observed in these three different models, and this might seems surprising given the different nature of the implemented medium effects. So it is worth asking what is  the most relevant common feature in these models, and one answer is the dominance of vacuum physics at early, high-energy stages of the shower~\cite{Du:2021pqa,Caucal:2018dla}. This brings in a key element for the interpretation:  large $\theta_g$ biases to more activity in the early vacuum shower. Since vacuum structures with more prongs lead to more quenched jets, the shape of $\theta_{g}$ is the consequence of a selection bias; high $\theta_{g}$ jets are more quenched and migrate to lower jet $p_{T}$ bins. Other models in the plot like the one denoted by Yuan et al indicate that flavour-dependent energy loss can also play a role.

Another substructure observable of interest is the $N$-subjettiness, denoted by $\tau_{N}$, which quantifies the degree to which a jet has a $N$(or fewer)-pronged substructure~\cite{Thaler:2010tr} 
%It is measured relative to $N$ axes, which are the axes of the subjets returned by unwinding the reclustering history of a given choice of reclustering algorithm by $N-1$ steps, and is defined as,                                                                                                      
 %\begin{equation}                                                                                                                                                              
%  \tau_{N}=\frac{1}{p_{\rm{T},\rm{jet}} \times R} %\sum_{k} p_{\mathrm{T},\it{k}} \rm{ } \: %\textrm{minimum} (\Delta \it{R}_{\mathrm{1},k},\rm{} %\Delta                         
%\it{R}_{\mathrm{2},k},...., \rm{} \Delta %\it{R}_{\mathrm{N},k}),                              
% \end{equation}                                                                                              %                                           where $k$ %runs over the list of jet constituents. The transverse %momentum, relative to the beam, of constituent $k$ is %denoted as $p_{\mathrm{T},k}$ and $\Delta              %R_{\mathrm{S},k}$ is the distance in the pseudorapidity-azimuthal ($\eta$-$\varphi$) plane between the constituent $k$ and the axis of subjet $S$. %The observable is           
%normalised by the product of the jet resolution parameter, $R$, and the jet transverse momentum, $p_{\rm T,\rm jet}$.       
 The ratio of $\tau_{2}/\tau_{1}$ is used to tag boosted hadronically-decaying objects such as the W and top quarks, which are typically 2-prong objects as compared to QCD jets, which are mostly 1-pronged. ALICE measured $\tau_{2}/\tau_{1}$\cite{Acharya:2021ibn} using several declustering metrics, including exclusive $k_{T}$ and CA+SD. The results do not reveal a significant change in the prong-structure of the jet relative to \pythia, which describes the observable well in pp collisions.

%\begin{figure}[ht]
%\centering
     
%      \includegraphics[width=0.4\linewidth]{plots/2018-09-29-Tau2to1_40to60_Full_Results_kT.pdf}
%      \includegraphics[width=0.4\linewidth]{plots/2018-09-29-Tau2to1_40to60_Full_Results.pdf}
%      \caption{Fully corrected $\tau_{2}/\tau_{1}$ distributions, measured with the $k_{\rm{T}}$(left), and C/A (right) with Soft Drop grooming algorithms, in Pb--Pb collisions at            
%$\sqrt{s_{\rm NN}} = 2.76$\,TeV for jets with $R = 0.4$ in                                                                                                                     
%the jet $p_{\rm T,\rm jet}^{\rm ch}$ interval of $40$--$60$\,GeV$/c$,}
      
%      \label{fig:tau2}
%\end{figure}
\begin{figure}[h]
\centering
       \includegraphics[width=0.45\linewidth]{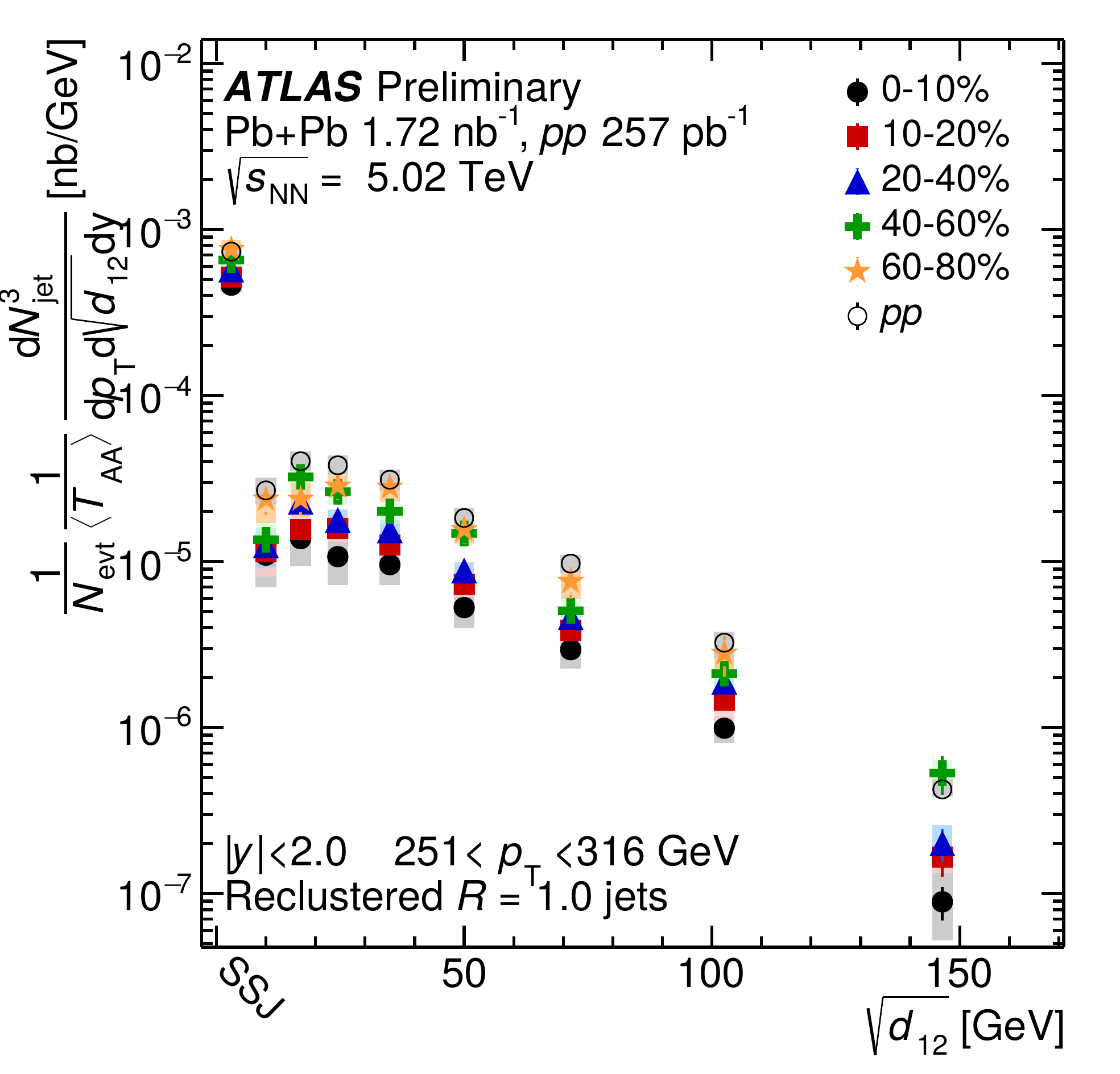}
      \includegraphics[width=0.45\linewidth]{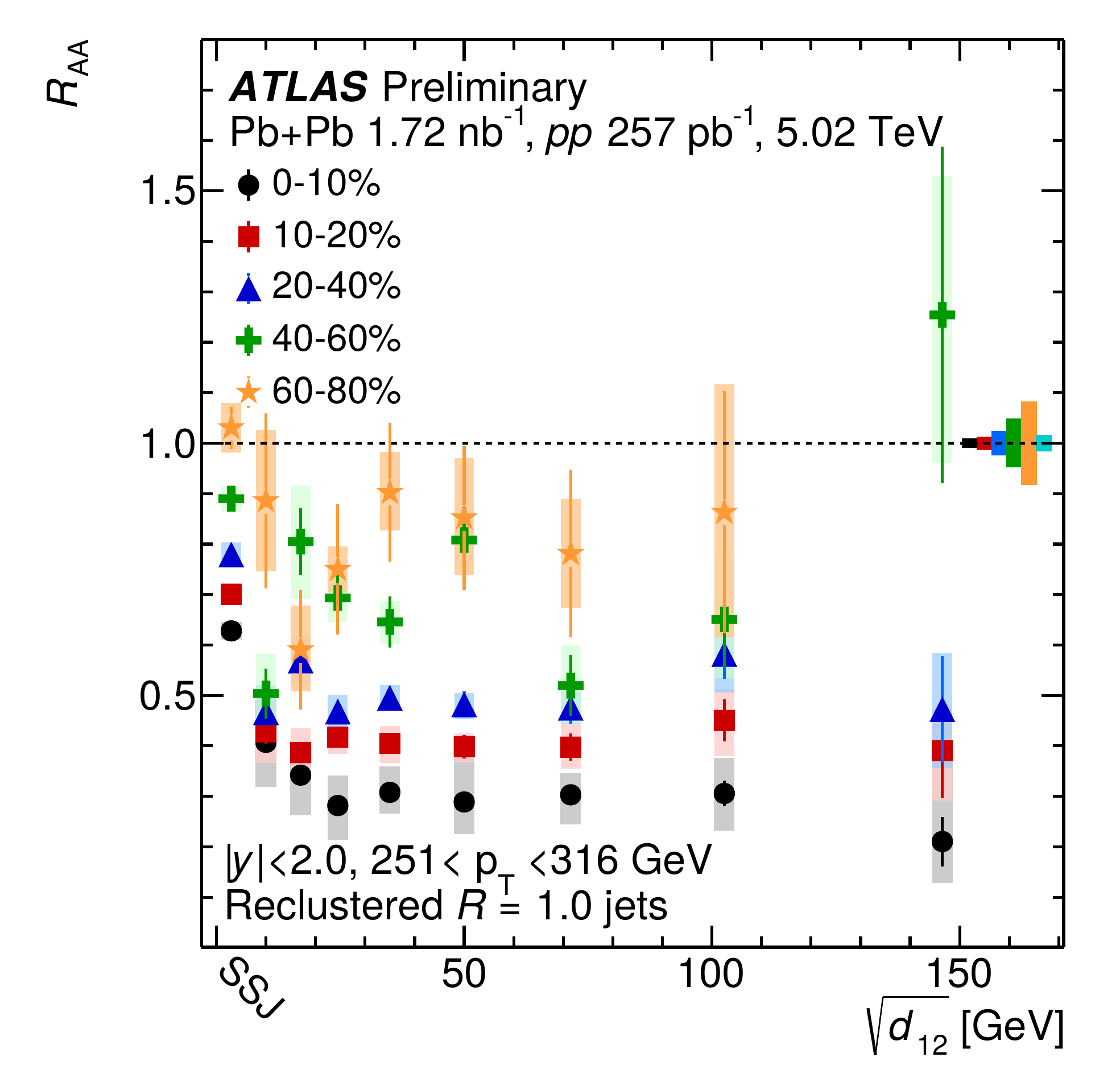}
     
      \caption{Left: Fully corrected $k_{T}$ distance measured for $R=1$ trimmed jets in the range $251<p_{T,jet}<316$ GeV for different centrality classes. Right: Nuclear modification factor as function of the $k_{T}$ distance. Figures are
      from Ref.~\cite{ATLAS:2019rmd}.}
      
      \label{fig:ktscale}
\end{figure}
ATLAS performed the first fully corrected measurement of the $k_{T}$ distance of large$-R$ jets in heavy ion collisions~\cite{ATLAS:2019rmd}. 
First, \RTwo\ calorimeter jets were reconstructed via the usual
procedure.  Then jets with $\ptjet > $~35~GeV
jets were taken as constituents for anti-$k_{T}$
jets clustered with $R = 1$. 
Their constituents were reclustered with the $k_{T}$ algorithm~\cite{Dokshitzer:1997in,Ellis:1993tq}  and then the last clustering step was unwound to register the $k_{\rm{T}}$ scale or distance, defined as 
\begin{equation} 
\sqrt{d_{12}}=min(p_{T,1}^2,p_{T,2}^2) \Delta R_{12}^{2} \end{equation} 
where indexes $1$ and $2$ refer to the two prongs that were clustered last.  
Large $\sqrt{d_{12}}$ selects jets with distinct hard prongs separated at large angles.
If an $R=1$ jet consists of only a single sub-jet (SSJ), $\sqrt{d_{12}}$ is not defined.
Figure~\ref{fig:ktscale} shows the $k_{T}$ distance distribution for different centralities and indicates that the majority of the jets consist of a 
single sub-jet. Two-prong configurations are suppressed by more than 2 orders of magnitude. 

The plot on the right shows the nuclear modification factor
is qualitatively different between those jets which have a single sub-jet
and those which have more than one.  Those jets with a single sub-jet are suppressed
approximately 50\% less in central collisions than those jets which have multiple
sub-jets.

In parallel to the writing of this review, other observables are being explored. An example is the subjet energy fraction, which considers the fraction of energy carried by the leading subjet within the signal jet. Another example is the transverse momentum $k_{\rm{T}}$ of the splitting found with dynamical grooming \cite{Mehtar-Tani:2019rrk}, $k_{\rm{T,dyn}}$, which selects the hardest splitting within the CA-ordered jet tree.

\begin{figure}[h!]
\centering
       \includegraphics[width=0.7\linewidth]{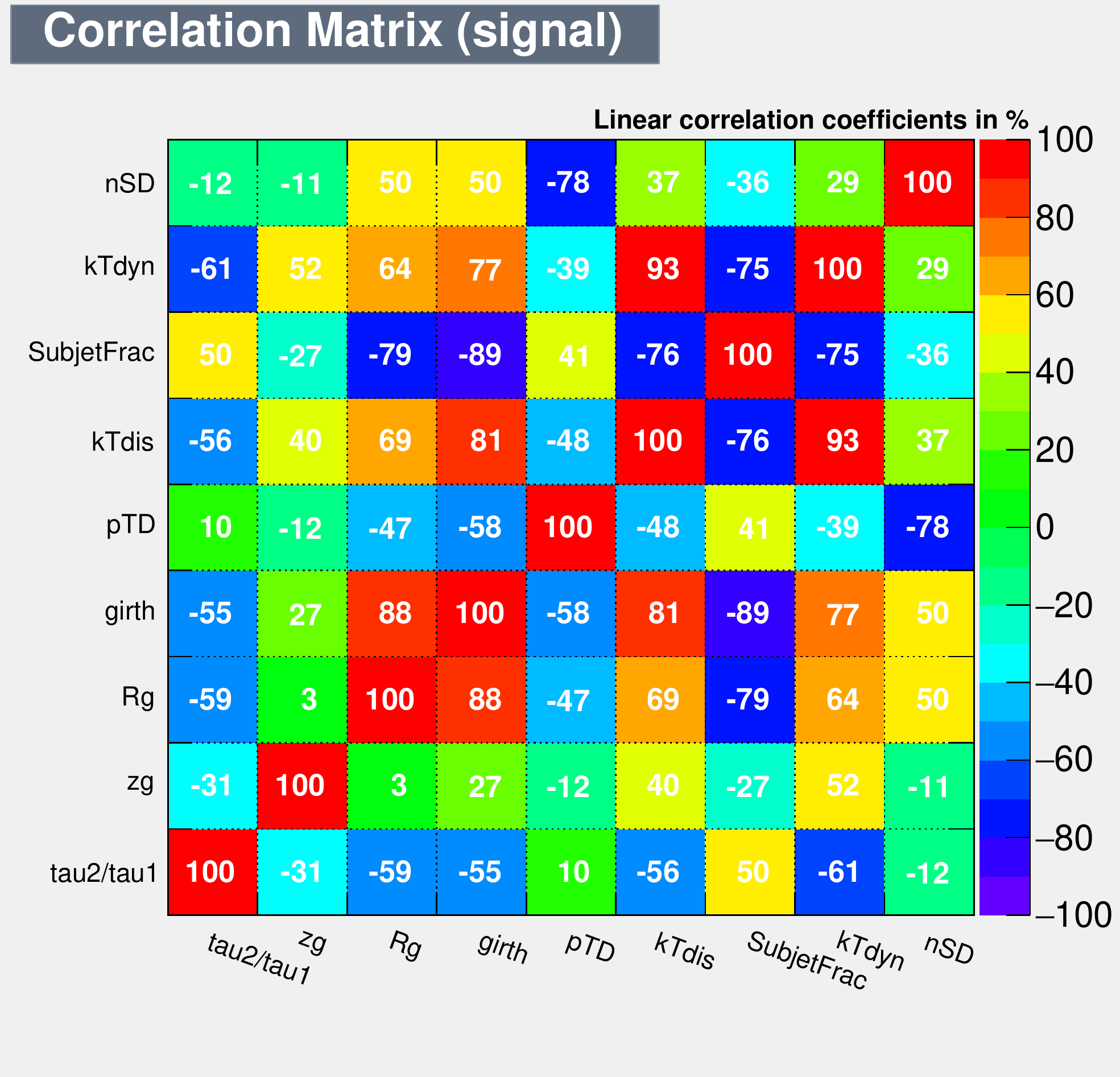}

      \caption{Linear Correlation matrix of the different jet shapes and substructure observables discussed in this chapter.}
      
      \label{fig:Correlations}
\end{figure}

All the discussed jet shape and  jet substructure observables must be correlated to some degree, by construction. For illustration, in Fig. \ref{fig:Correlations} we show the linear correlation coefficients for PYTHIA8  \cite{Sjostrand:2007gs} jets reconstructed with $R=0.4$ with \ptjet\ $>100$ GeV. We observe that the $k_{\rm{T}}$ distance is strongly correlated to the girth and to $k_{\rm{T,dyn}}$ and strongly anti-correlated to the leading subjet fraction. The $n_{\rm{SD}}$, which is a measure of the intrajet multiplicity is naturally anti-correlated to the $p_{\rm{TD}}$ which is related to the dispersion in momentum of the jet constituents. 
The $z_{g}$ measures a momentum balance while $R_{g}$ is an angle and they are not correlated. We also note the strong correlation between the girth and $R_{g}$, $k_{\rm{T,dyn}}$, and the $k_{\rm{T}}$ distance. 
Finding a set of minimally correlated observables can be useful to perform systematic comparisons to models and calculations. An example of such procedure in \pp\ collisions is the recent extraction of $\alpha_{S}$ using jet substructure in $t\bar{t}$ events by CMS, where $R_{g}$,$z_{g}$,$\epsilon$ and jet multiplicity were identified as a set of minimally correlated variables among more than $30$ substructure observables \cite{CMS:2018ypj}.   

The selection bias was discussed in the context of the $R_{g}$ but applies to most of the discussed observables. In order to mitigate this selection bias, and to increase the weight of quenched jets in the measured samples, different strategies are envisaged. An obvious one considers the substructure of jets recoiling from Z or $\gamma$ bosons. Other interesting approaches based on ML have been proposed~\cite{Du:2021pqa}.

\subsubsection{What happens to the energy lost from jets in the QGP?}

The main physics aim here is to understand the process by which 
energy lost by the jet is incorporated into the QGP.  There are 
two reasons why this is important:
\begin{itemize}
\item this provides access to the hydrodynamization process 
\item the energy from the medium response is correlated with the 
jet and affects other observables which are used to quantify the strength
of energy loss~\cite{He:2018xjv,Pablos:2019ngg}.
\end{itemize}

Three kinds of observables have been used to search for this effect:
\begin{itemize}
\item cone size dependence of jet \RAA
\item correlations between jets and tracks
\item fragmentation functions and jet shapes

\end{itemize}

%Measurements of fragmentation functions and jet shapes provide information on 
%how the particles inside the jet carry the momentum.  In these measurements 
%ideally the measurement of the jet itself is independent of the measurement of the
%constituents, providing an unbiased measurement of the constituents.  

%Experimentally, these measurements are straightforward to measure both in heavy-ion
%collisions and \pp\ collisions.  Measurements in \pp\ collisions provide a baseline
%for unmodified jets.  Theoretically, these measurements can be compared to generator
%results (such as pythia and herwig) as well as model calculations.

In order to capture the full dynamics of jet quenching, large-$R$ jets and access to their internal structure is desired. Cross sections and the ratios of cross sections for different $R$ are IRC-safe observables that can be analytically calculated and pose strong constrains to the theory. 
The heavy-ion underlying event creates combinatorial or fake jets that prevent unfolding and only at very high jet $p_{\rm{T}}$ the measurement of the inclusive large-$R$ jet is feasible. Below 100 GeV, the different collaborations have measured
jet cross sections and their ratios for different $R$ up to $R=0.5$~\cite{ATLAS:2012tjt,ALICE:2015mdb}.

\begin{figure}[h!]
 \centering
 \includegraphics[width=0.9\textwidth]{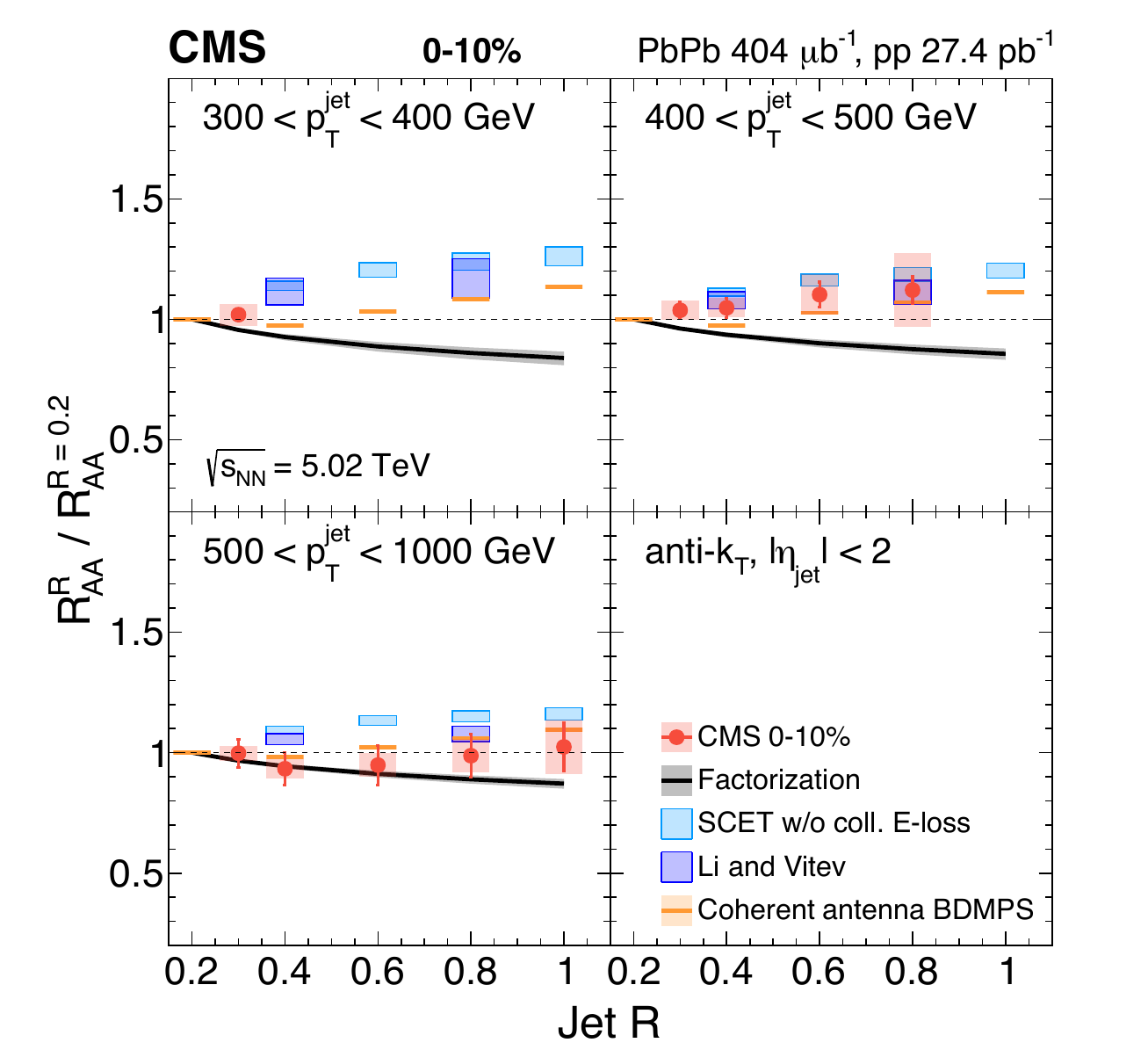}
 \caption{ Ratio of the $R_{\rm{AA}}$ for different jet $R$ and $R_{\rm{AA}}$ for $R=0.2$, in different jet $p_{\rm{T}}$ intervals and compared to several calculations. Fig. from Ref.~\cite{CMS:2021vui}.}
 \label{fig:CMSR}
 \end{figure}

In the energy range of a few hundred GeV up to the 1 TeV, CMS has reported the first measurement of jet nuclear modification factors for jets with radii from $R=0.2$ up to $R=1$, for different centrality classes. 
In central collisions, and up to $R=0.4$ (where there are still sufficient data points to observe a trend), the nuclear modification factor increases with jet $p_{\rm{T}}$, in agreement with the ATLAS result for $R=0.4$ jets~\cite{ATLAS:2018gwx}. Above 500~GeV, where a full scan of the $R$ dependence is possible, the data is consistent with no dependence of the \RAA with jet $R$. The comparison of the data to models and calculations reveals significant tensions in the simultaneous description of the nuclear modification factor and its $R$ dependence. In Fig. \ref{fig:CMSR} an example of comparisons to analytical calculations is shown.

What is common to many of the models compared to the CMS data, is the strong role of the medium response which gives a larger contribution at large $R$. In models like the Hybrid model, the $R$ dependence of $R_{\rm{AA}}$ can be explained as the result of the balance of a stronger suppression for broader jets, and the ability to include more
medium response inside the cone.
The high transverse momentum of the $R=1.0$ jets in the CMS measurement could limit the 
effect of medium response.  Measurements with a larger kinematic 
range will allow for better discrimination between models.

Other measurement that emphasizes the role of the medium response at large $R$ is for instance the jet mass \cite{ALICE:2017nij} for $R=0.4$ jets. No modifications in \pbpb\ collisions relative to \pPb\ collisions
were observed, possibly due to a balance of energy loss and medium response effects~\cite{KunnawalkamElayavalli:2017hxo}.

%\begin{figure}[h!]
 %\centering
 %\includegraphics[width=0.9\textwidth]{plots/CMS%-HIN-18-014_Figure_012.png}
 %\caption{From Ref.~\cite{Sirunyan:2021pcp}. To %choose between 14 and 15. I think this one %relates better to the text, but no strong %opinion}
 %\label{fig:CMSRb}
 %\end{figure}

 In order to look for medium response, measurements
 of low momentum tracks around jets in heavy-ion collisions have been
 performed.
There has been interest in measuring fragmentation
functions as a function of \ptpart, the 
transverse momentum of the particle in the jet.  This 
is motivated to search for an
absolute scale in the modification of the fragmentation.  
When looking at the jets fragmentation functions plotted as a function
of \ptpart, the low-\ptpart, $\ptpart\ <$~4~GeV, part of these ratios are approximately equal
for the three jet \pt\ selections.  
The low-\ptpart\ excess is thought to be due to the response of the
medium to the passage of the jet.  The approximate scaling and extent in \ptpart\ 
of the excess would be then sensitive to some scale in the QGP associated with the 
response. 
%The approximate $\ptpart\ <$~4~GeV scale is something that will show up in
%other measurements of the particles within jets discussed below.
 
\begin{figure}[h!]
\centering
\includegraphics[width=0.43\textwidth]{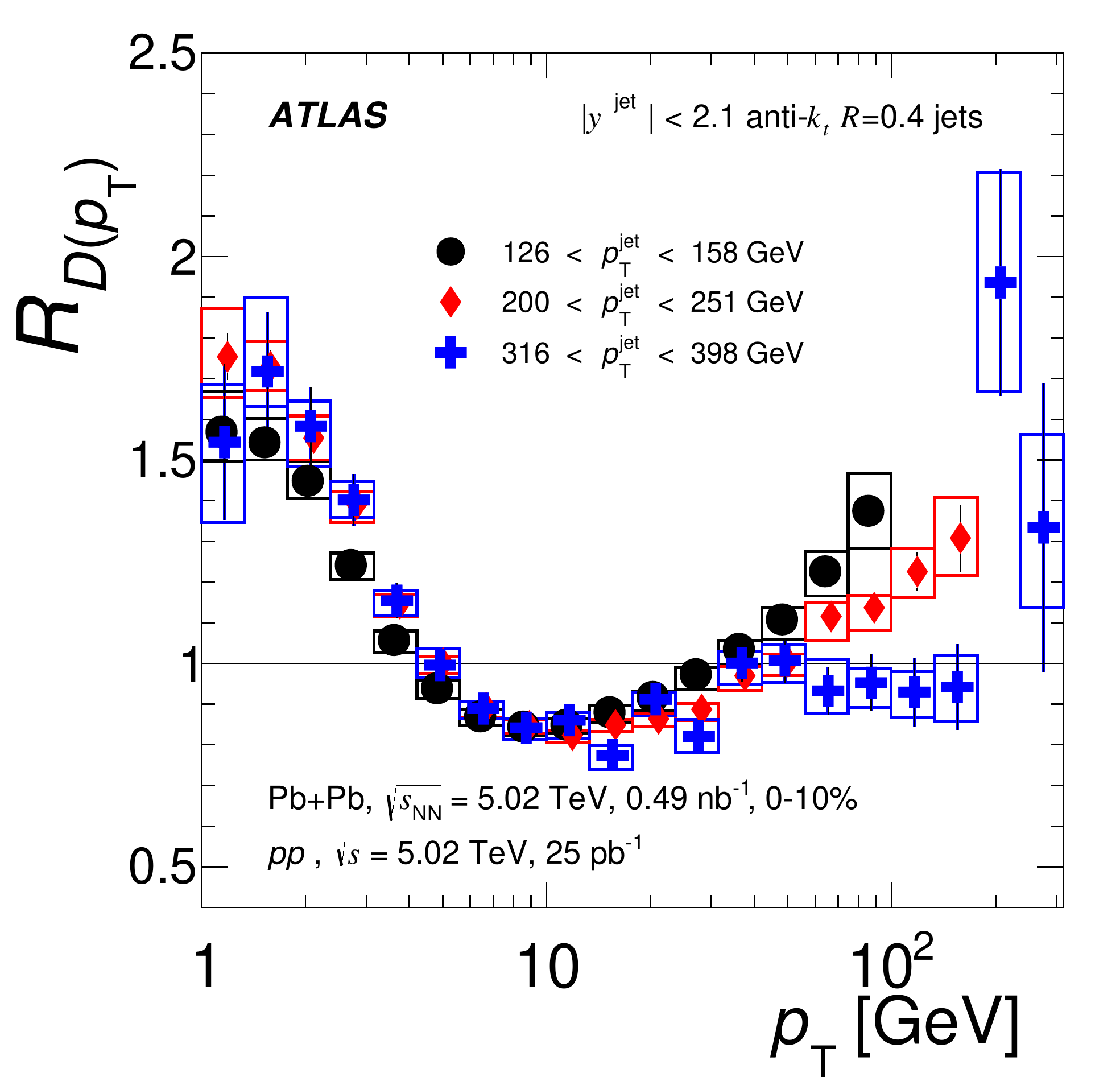}
\includegraphics[width=0.55\textwidth]{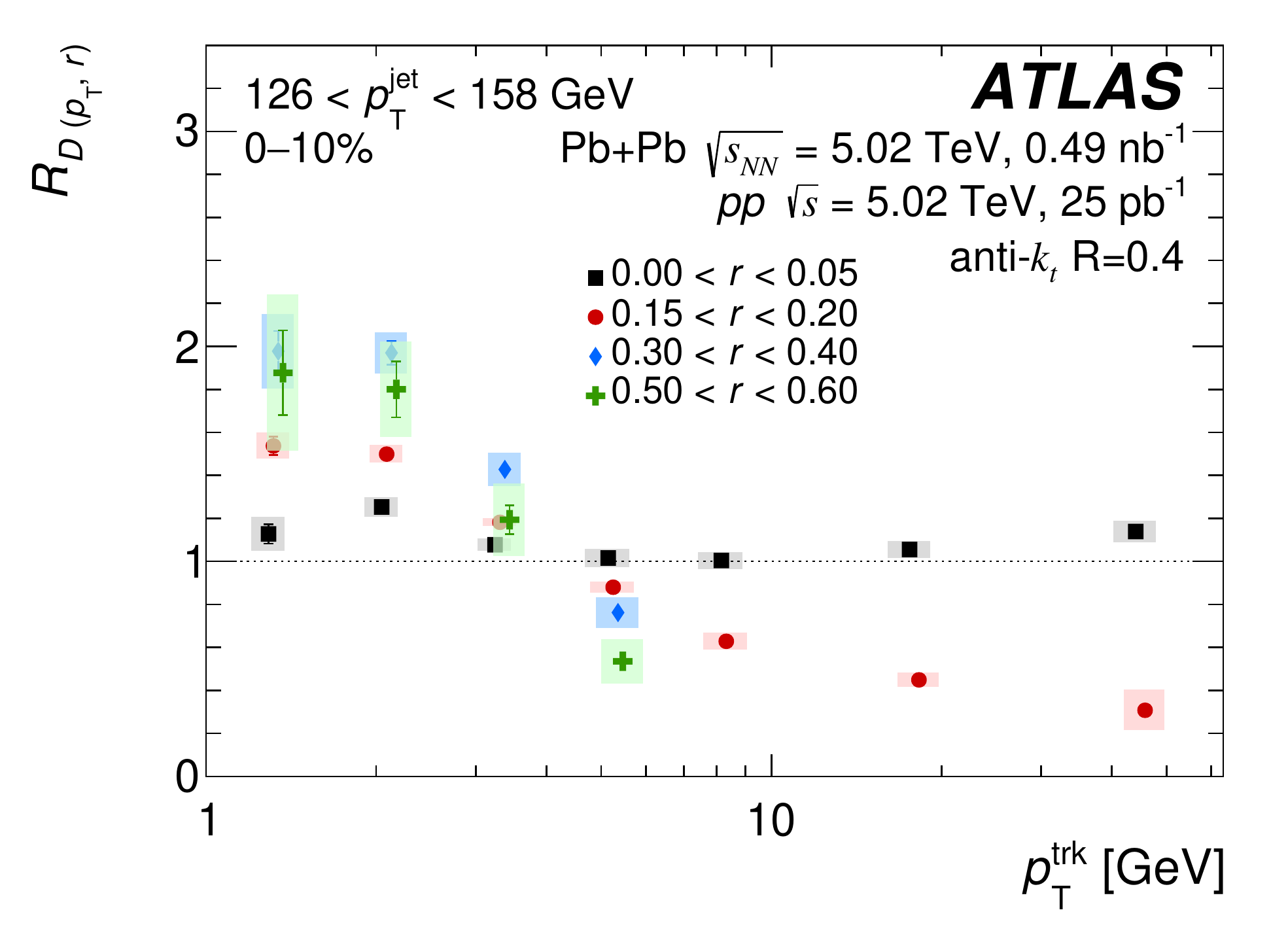}
\caption{Left: Ratios of the fragmentation functions in 
central \pbpb\ collisions to those in \pp\ collisions
for three different \ptjet\ selections as a function of charged-particle \pT.
Right: The same quantity as in the left plot only differential in the distance, \r,
from the jet axis as well (different sets of points).
Figures are from Ref.~\cite{Aaboud:2018hpb} (left) and ~\cite{Aad:2019igg}
(right).}
\label{fig:FF_PbPb}
\end{figure}

 In order to study both the angular and longitudinal directions at once,
 both CMS and ATLAS have measured two-dimensional fragmentation 
 functions~\cite{Khachatryan:2016erx,Khachatryan:2016tfj,Sirunyan:2018jqr,Aad:2019igg,Sirunyan:2021jty}.
 In the longitudinal fragmentation functions we noted that the low-\ptpart\
 excess was for particles below approximately 4 GeV.  The two-dimensional
 fragmentation functions in Refs.~\cite{Khachatryan:2016erx,Sirunyan:2018jqr,Aad:2019igg}
 provide support for that approximate scale both in 2.76 and 5.02 TeV \pbpb\ collisions
 at the LHC.  Figure~\ref{fig:FF_PbPb} shows the ratio of the two-dimensional
 fragmentation function in central \pbpb\ collisions to that in \pp\ collisions
 as a function of \ptpart\ for different values of distance $r$ to the jet axis.
 The magnitude of the modifications changes as a function of $r$, but the location 
 in \ptpart\ of the transition from suppression to enhancement is at approximately
 4 GeV for all $r$ values. 
 This same 4~GeV scale is also seen in measurements of $Z$-hadron
 correlations at ATLAS~\cite{ATLAS:2020wmg}.

\subsection{Effective Degrees of Freedom of the QGP}
The QGP behaves macroscopically as an almost perfect liquid. However, since the underlying theory is Quantum Chromodynamics, it is expected that if the QGP is probed at sufficiently short distances, the   quasi-particle degrees of freedom will emerge \cite{DEramo:2012uzl}. 
If the QGP were strongly coupled at all scales, the distribution of transverse momentum $k_{T}$ transferred from the medium to an energetic quark or gluon projectile is expected to be Gaussian. In the limit where the parton projectile resolves the free quarks and gluons within the QGP, the distribution of transferred transverse momentum is expected to follow a power-law tail $1/k_{T,4}$, typical of point-like scatterers. This is often referred to as the Moliére regime. 

The searches for point-like scatterers in the QGP can be done both at inter-jet and intra-jet level. In the inter-jet case, the azimuthal correlation between a high-pt hadron, or ideally a photon or a boson and the recoiling jet is measured
and compared the yield of large-angle deflections in \pbpb\ and \pp\ collisions in the search for an excess.  The intra-jet case utilizes new substructure techniques in order to identify high-$k_{T}$ prongs or splittings within the jet. An example of such techniques is the dynamical grooming \cite{Mehtar-Tani:2020oux}, which allows to select the hardest prong in the jet tree. New NLO calculations of the medium-induced radiative spectrum within the Improved Opacity Expansion \cite{Barata:2021wuf} and their ongoing extension to substructure will provide analytical reference to the expected impact of the power-law tail. 

\begin{figure}[h!]
\centering
\includegraphics[width=0.45\textwidth]{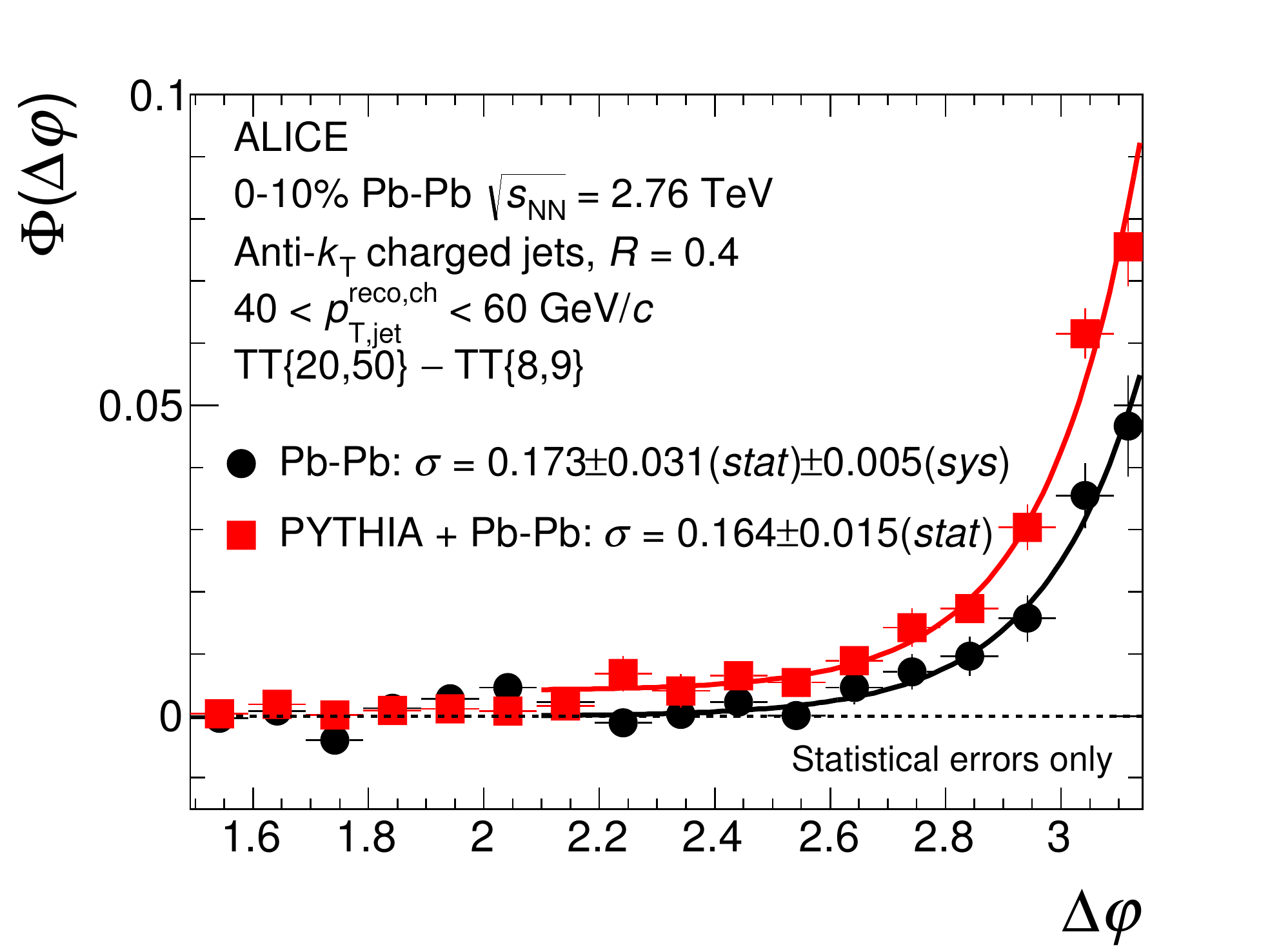}
\includegraphics[width=0.45\textwidth]{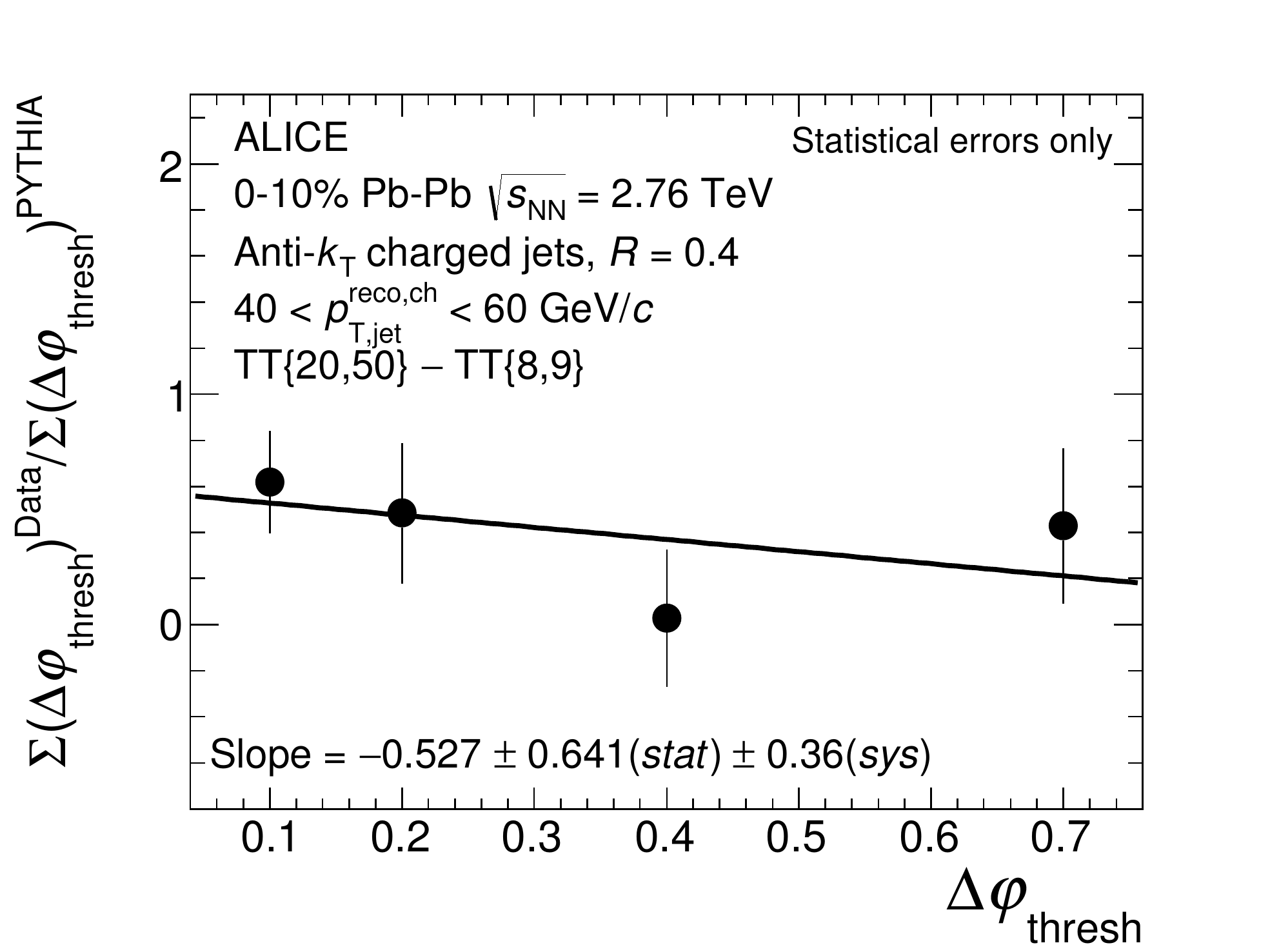}

\caption{From Ref. \cite{ALICE:2015mdb} Left: Azimuthal correlation between a high-$p_{T}$ hadron and the recoiling jets in Pb-Pb collisions and in the vacuum PYTHIA calculation. Right: Large-angle deflections are examined by integrating the yield of the azimuthal correlation from $\pi/2$ to $\pi-\varphi_{thresh}$ } 
\label{fig:LargeAngleALICE}
\end{figure}

\begin{figure}
\centering
\includegraphics[width=0.9\textwidth]{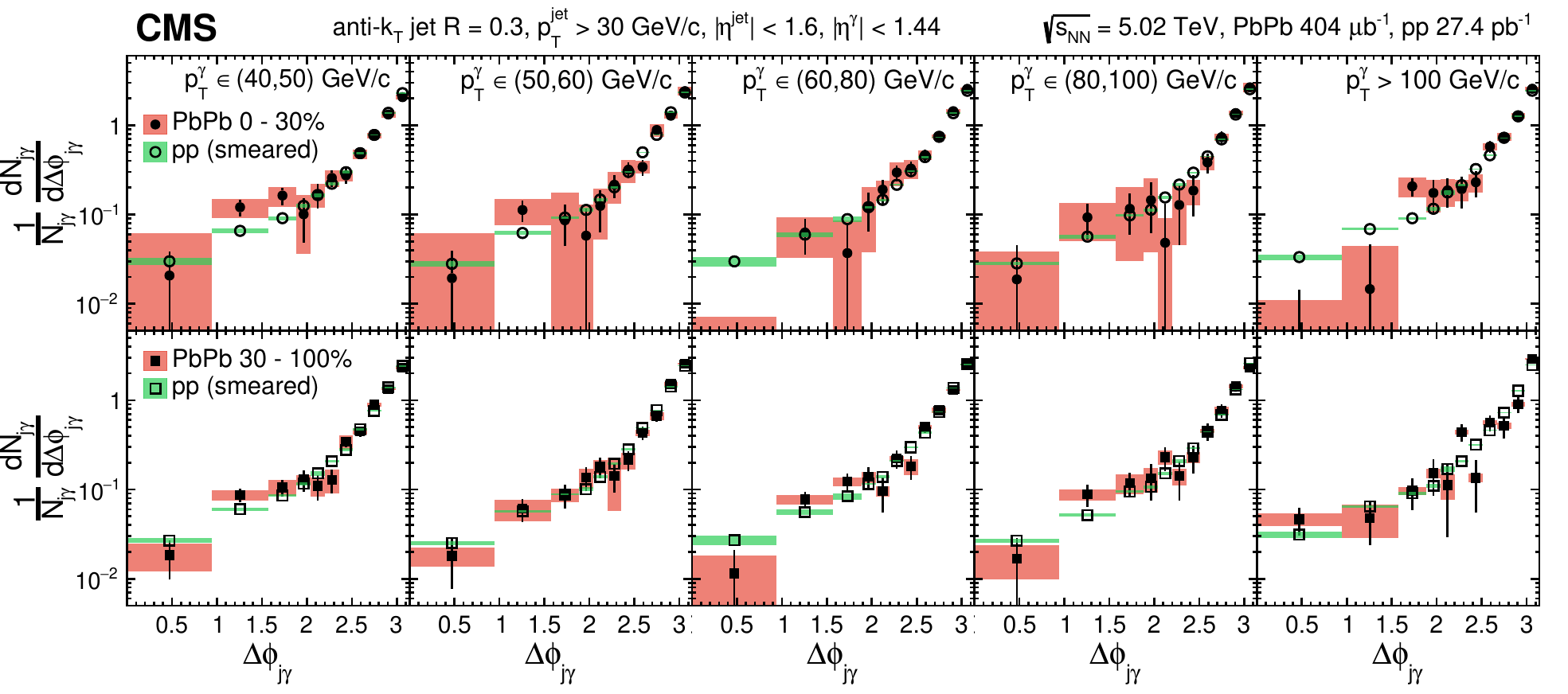}
\caption{CMS gamma-jet azimuthal correlation in PbPb compared to \pp\ collisions from Ref. \cite{CMS:2017ehl}} 
\label{fig:AzimuthalCMS}
\end{figure}

Inter-jet azimuthal correlations have been studied by ATLAS~\cite{ATLAS:2010isq}, CMS \cite{CMS:2017ehl}, ALICE~\cite{ALICE:2015mdb} and STAR~\cite{STAR:2017hhs} collaborations in different kinematic regimes. As an example, the ALICE semi-inclusive azimuthal correlation between high-momentum hadrons and jets is shown in Figure \ref{fig:LargeAngleALICE}, together with the accumulated integrated yield on the right plot. The statistical precision of the data doesn't allow conclusions to be drawn on a possible modification of the yield at very large angles. However the ALICE measurements sketches what can be done in the near future with higher statistics, with a \pp\ reference instead of a MC calculation and with a full and symultaneous correction of the recoil jet momentum and the azimuthal angle and a full kinematic scan of the trigger object (ideally a photon) and the recoil jet $p_{T}$. The CMS measurement of the azimuthal correlation between isolated photons and jets is shown in Figure~\ref{fig:AzimuthalCMS}. The sensitivity to large-angle modifications is limited by statistics  and by the systematics of the uncorrelated background subtraction. 

This section focuses on the tails of the azimuthal correlation, but the bulk of the correlation is also of interest to probe broadening due to multiple soft scatterings with the medium. Calculations have shown that the sensitivity to medium effects is enhanced at low photon and recoil jet energies, since at high energies the distribution is dominated by vacuum radiation~\cite{Mueller:2016gko} and medium effects are indistinguishable.

\subsection{Critical Size for QGP Formation}
\subsubsection{Searches for Jet Quenching in \pA\ collisions}

Given the recent wealth of data showing evidence for 
collective behavior in
\pPb, \pp\ collisions (for a recent review see Ref.~\cite{Nagle:2018nvi}), 
it is natural to search for
jet quenching effects in these systems.  As of this writing
no effects of jet quenching have been observed in \pPb\ or \pp\ collisions.
Here we discuss several searches for jet quenching in \pA\ collisions.

The nuclear modification factor \RpA\ has been measured both in
\dAu\ collisions at RHIC and \pPb\ collisions at the LHC
for both jets~\cite{Adare:2015gla,Adam:2015hoa,ATLAS:2014cpa} and 
charged particles~\cite{Adams:2003im,Adler:2003ii,Khachatryan:2016odn}.  
No evidence for jet quenching was found in these measurements.
However, the precision of these measurements is limited by the
normalization uncertainties associated with the nuclear modification
factor coming from the luminosity and \TAA\ determination (along 
with other sources).

\begin{figure}[h!]
 \centering
  \includegraphics[width=0.45\textwidth]{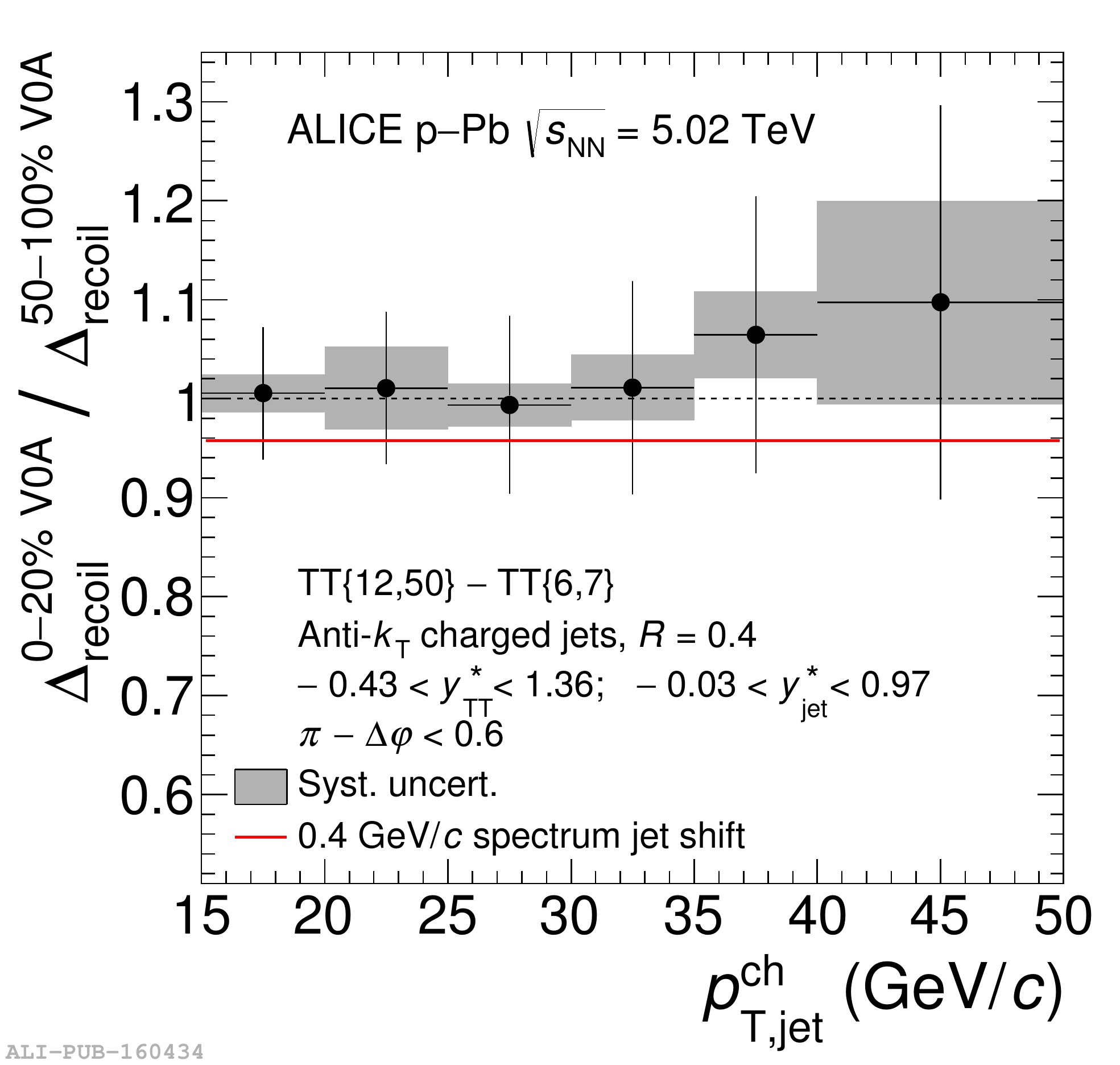}
 \caption{ Ratio of recoil jet $p_{\rm{T}}$ distribution in \pPb\ events with high and low event activity measured in the forward detectors. From Ref. \cite{Aaboud:2017tke}}
 \label{fig:pPbjqALICE}
 \end{figure}

In order to be sensitive to potentially smaller jet quenching 
effects, measurements of self-normalized observables (e.g. normalized per-jet, or per-trigger
particle) have been done.
ALICE measured the charged-particle jets opposite to a high-transverse momentum 
trigger hadron~\cite{Acharya:2017okq} and reported their per-trigger normalized yield over a broad kinematic range.
The advantage of the per trigger particle normalisation is that no \TAA\ scaling of the reference is needed and thus no Glauber modelling and interpretation of the event activity (EA) in terms of geometry is required. 
Events are classified according to how hits in a forward scintillator in the Pb-going
direction or hits in a zero-degree neutron detector also in the Pb-going direction. 
Figure \ref{fig:pPbjqALICE}, shows the ratio of the observable in the two EA classes. The ratio is consistent with no energy loss. The red line indicates a limit, at $90\%$ confidence level, on the average $p_{T}$ shift of 0.4 GeV/c, which is an estimate of the maximum energy that is transported outside the jet cone.

\begin{figure}[h!]
 \centering
 \includegraphics[width=0.59\textwidth]{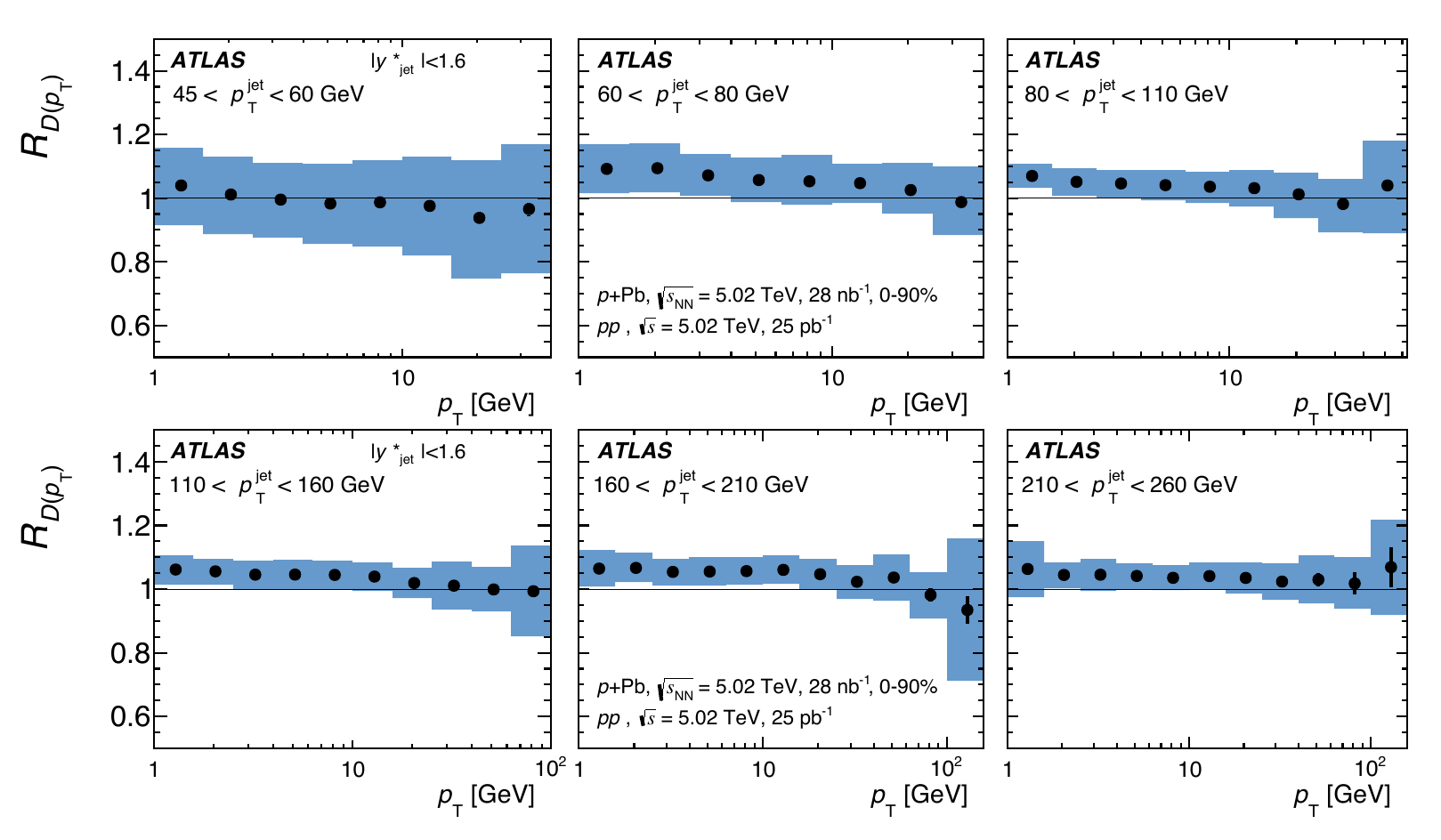}
 \includegraphics[width=0.4\textwidth]{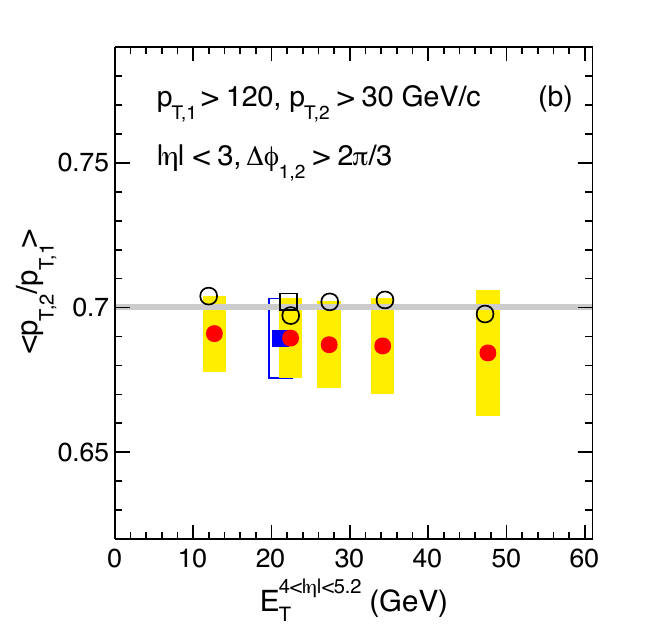}
 \caption{(left) Ratios of the fragmentation functions in \pPb\ collisions
 to those in \pp\ collisions for various \ptjet\ selections as a
 function of charged particle \pT.  (right) Mean value
 of the ratio of the subleading to leading \ptjet (\xJ) 
 as a function of the transverse energy in the Pb-going direction.  Figures
 are from Ref.~\cite{Aaboud:2017tke} (left) and Ref.~\cite{Chatrchyan:2014hqa} (right). }
 \label{fig:pPbjq}
 \end{figure}

 ATLAS measured fragmentation functions in \pPb\ collisions~\cite{Aaboud:2017tke}.  
 As shown in Figure~\ref{fig:pPbjq} in contrast
 to \pbpb\ collisions, no significant deviation was found between the \pPb\ fragmentation
 functions and the \pp\ ones for the soft particles in the jet.  
 There is some excess of less than 10\% in the central values 
 for charged-particles between approximately 1--5~GeV between 
 the fragmentation functions in \pPb\ and \pp\ collisions but it is within the
 size of the systematic uncertainties. 
 Also shown in Figure~\ref{fig:pPbjq} is the measurement from CMS of 
 the dijet asymmetry in \pPb\ collisions selected on the forward
 energy in the Pb-going direction~\cite{Chatrchyan:2014hqa}.  
 They evaluated the mean of this distribution
 and found that quantity to be independent of the forward energy to within their
 uncertainties.
 
There is ongoing interest in more sensitive measurements which might
be more sensitive to any jet quenching signal but the existing measurements
clearly show that any jet quenching which might exist in \pPb\ collisions
is much smaller than that in heavy ion collisions.

\subsubsection{Light ion collisions}
 In order to map the transition
between large systems (e.g. central heavy ion collisions collisions) with large
energy loss and small \pA\ systems without observed energy
loss, there is a lot of interest in having small, symmetric 
collision systems with which to potentially observe the turn-off of jet
quenching.

Some data from collision systems smaller than \pbpb\ or
\auau\ does exist.  Most recently, the LHC delivered
\xexe\ collisions in 2017.  Those results showed that the
value of \RAA\ for charged particles depends primarily
on the size of the collision system (as measured by the charged particle
multiplicity)~\cite{Acharya:2018eaq}.
However, the utility of the \xexe\ measurements to answer this question
is limited by the 
fact that \xexe\ collisions are much closer in charged-particle
multiplicity and \Npart\ to \pbpb\ collisions than they are to \pPb\
collisions.

In light of this there remains a great interest in colliding a 
much smaller collision system, with an \Npart\ close to that 
of \pPb\ collisions but with a larger geometrical transverse overlap that increases the in-medium path length and thus potentially, quenching. 
The preferred collision system
is \OO~\cite{Huss:2020dwe,Brewer:2021kiv}. The experimental projections for the nuclear modification factor of charged hadrons measured in a short \OO\ run of $0.5$ nb$^{-1}$ \cite{ALICE-PUBLIC-2021-004} were compared to theoretical expectations for jet quenching ~\cite{Huss:2020dwe,Huss:2020whe}. The comparison indicates that partonic energy loss signal might be observed at transverse momentum of approximately 20~GeV. 

In the spring of 2021
RHIC ran \OO\ collisions for the STAR experiment.
The data from that run has not yet been analyzed but will 
be the first look at this important question.

\section{Conclusions and Outlook}

This review covers the current status of measurements which 
use jets to study the properties of the quark-gluon plasma.  The 
initial observations of jet quenching at RHIC and later at the LHC
were only the beginning of this rich program.  

Measurements
covered here clearly show that the amount of energy loss a jet
undergoes depends on the structure of the jet. Broader jets and jets with a distinct two-prong substructure with a large opening angle appear to be more suppressed than inclusive jets.  This observation
clearly motivates further measurements sensitive to the jet
flavor and substructure as essential to understanding the
interactions between jets and the QGP.  Among those measurements, the substructure of jets recoiling from a photon or Z-boson are of particular interest to mitigate the selection bias present in inclusive measurements.

Measurements have also
shown that there is a wide distribution of soft particles 
around jets.  This is thought to be from the response of the QGP
to the jet passing through it.  This can provide further information
about the transport of energy in the QGP. Several of the measurements discussed in this review can be described as a balance between energy loss, and the recovery of the medium response to the jet at large angles. 

Finally, we have discussed
measurements which show how jet quenching depends on geometry. 
More measurements sensitive to this, including the effects of
fluctuations in the initial geometry of the collision are essential 
to understand the path length dependence of energy loss. 
%the other questions in energy loss because all jets are measured having traversed
%a unique QGP profile.

The limit of jet quenching in very small QGP systems is an area
of great interest.  Evidence for collectivity  is well established in
\pPb\ collisions, however jet quenching has not been observed. 
%Azimuthal anisotropies at high-\pT\ asssociated with jet quenching in \pbpb\ collisions
%are observed in \pPb\ collisions but without the suppression in the number
%of jets compared to \pp\ collisions.  
Reconciling these measurements into a 
common interpretation is a key physics aim of the light-ion program
underway at RHIC and expected in Run 3 (2024) at the LHC.

Looking forward we anticipate a few exciting opportunities in this field in
the next few years.  First, there are exciting experimental and theoretical
investigations into jet substructure ongoing.  These will be key in understanding
how energy loss happens.  Second there will be a wealth of new data, 
including Run 3 at the LHC with increased luminosity and the first data taken with 
the ALICE upgrades and the turn on of
sPHENIX at RHIC.  sPHENIX will provide fully calorimetric jets at RHIC for 
the first time and will have a data recording rate which will allow
for a greatly expanded kinematic range of jets at RHIC.  This will allow
for new constraints on the dependence of jet quenching on the QGP temperature.

\section{Acknowledgements}
The authors thank Matthew Nguyen, Martin Rybar, Carlos
Salgado, and Marco van Leeuwen 
for comments and suggestions to the draft. The authors also thank the ALICE, ATLAS, CMS, PHENIX and STAR Collaborations for the great experimental results.
LCM is supported by the European Research Council project ERC-2020-COG-101002207 QCDHighDensityCMS.
AMS acknowledges support from  National Science Foundation
Award Number 2111046.

%\addcontentsline{toc}{chapter}{References}
\section{References}
\bibliographystyle{unsrturl}
\bibliography{refs}

\end{document}